\documentclass[preprint, 
 amsmath,amssymb,
 aps, physrev,
]{revtex4-2}

\usepackage{graphicx}\usepackage{dcolumn}\usepackage{bm}\usepackage{subcaption}

\usepackage[T1]{fontenc}
\usepackage[utf8]{inputenc}
\usepackage[a4paper]{geometry}
\usepackage[table, dvipsnames]{xcolor}
\usepackage{amsmath, amssymb, amsthm}
\usepackage{afterpage}
\usepackage{xpatch, xparse}
\usepackage{booktabs}
\usepackage{array}
\usepackage{caption}
\usepackage{tikz}
\usepackage{textpos}
\usepackage{grffile}

\usepackage[pdfpagelabels=false]{hyperref}	\hypersetup{colorlinks=true,linkcolor=blue,citecolor=blue,filecolor=blue,urlcolor=blue,}

\usepackage{cleveref}

\newcommand{\progname}[1]{{\sc #1}}

\begin{document}

\preprint{APS/123-QED}

\title{\textbf{FCC feasibility studies: Impact of tracker- and calorimeter-detector performance on jet flavor identification and Higgs physics analyses} 
}

\author{Haider Abidi} \affiliation{Physics Department, Brookhaven National Laboratory, Upton NY; United States of America.}
\author{Ketevi A. Assamagan} \affiliation{Physics Department, Brookhaven National Laboratory, Upton NY; United States of America.}
\author{Diallo Boye} \affiliation{Physics Department, Brookhaven National Laboratory, Upton NY; United States of America.}
\author{Elizabeth Brost} \affiliation{Physics Department, Brookhaven National Laboratory, Upton NY; United States of America.}
\author{Viviana Cavaliere} \affiliation{Physics Department, Brookhaven National Laboratory, Upton NY; United States of America.}
\author{Anna E. Connelly} \affiliation{Physics Department, Brookhaven National Laboratory, Upton NY; United States of America.}
\author{George Iakovidis} \affiliation{Physics Department, Brookhaven National Laboratory, Upton NY; United States of America.}
\author{Ang Li} \affiliation{Physics Department, Brookhaven National Laboratory, Upton NY; United States of America.}
\author{Marc-Andr\'e Pleier} \affiliation{Physics Department, Brookhaven National Laboratory, Upton NY; United States of America.}
\author{Andrea Sciandra} \email{andrea.sciandra@cern.ch} \affiliation{Physics Department, Brookhaven National Laboratory, Upton NY; United States of America.}
\author{Michele Selvaggi} \affiliation{CERN, Geneva; Switzerland}
\author{Scott Snyder} \affiliation{Physics Department, Brookhaven National Laboratory, Upton NY; United States of America.}
\author{Robert Szafron} \affiliation{Physics Department, Brookhaven National Laboratory, Upton NY; United States of America.}
\author{Abraham Tishelman-Charny} \affiliation{Physics Department, Brookhaven National Laboratory, Upton NY; United States of America.}
\author{Iza Veliscek} \affiliation{Physics Department, Brookhaven National Laboratory, Upton NY; United States of America.}

\date{\today}

\begin{abstract}
The extensive and ambitious physics program planned at the Future Circular Collider for electrons and positrons (FCC-ee) imposes strict constraints on detector performance. This work investigates how different detector properties impact jet flavor identification and their subsequent effects on high-profile physics analyses. Using Higgs boson coupling measurements and searches for invisible Higgs decays as benchmarks, we systematically evaluate the sensitivity of these analyses to tracker and calorimeter detector configurations. We examine variations in single-point resolution, material budget, silicon layer placement, and particle identification capabilities, quantifying their effects on flavor-tagging performance. Additionally, we present the first comprehensive study of Higgs-to-invisible decay detection using full detector simulation, providing important insights for optimizing future detector designs at lepton colliders.
\end{abstract}

\maketitle

\section{Introduction}

The Future Circular Collider for electrons and positrons (FCC-ee) project~\cite{aut2018} aims to explore particle physics frontiers with unprecedented precision. Several detectors have been proposed~\cite{IDEA1,IDEA2,cld,ALLEGRO}, each with specific sub-detector configurations and geometries. This study examines how different detector choices impact high-profile analyses, particularly focusing on tracker configurations and their effects on jet flavor identification (jet-tagging) and variations of the calorimeter properties. We evaluate how these choices influence Higgs-boson coupling measurements and Higgs self-coupling analysis. Additionally, we assess various calorimeter properties for detecting invisible Higgs-boson decays.

Accurate jet-flavor identification is crucial for measuring Higgs-boson couplings. These algorithms distinguish between fundamental particles produced in high-energy collisions, enabling precise Standard Model measurements and detection of new physics phenomena. They are essential for the FCC-ee physics program's core objectives: precision Higgs-boson studies, top-quark measurements, and searches for physics beyond the Standard Model. These algorithms guide detector optimization and drive technological advancement.

An algorithm~\cite{FCCFlavTag} for reconstructing track parameters and covariance matrices of charged particles for arbitrary tracking geometries was developed, alongside a graph neural network-based jet-flavor identification system. A generalization of this neural network approach to transformer networks was performed, see Ref.~\cite{GenTransformers} and~\cite{SarasFCCNote}, and is adopted in the following studies.
Using the FCC-ee IDEA detector prototype, we evaluated various configurations: adding a fourth innermost pixel detector layer, and removing particle-identification data through either or both the cluster-counting ($dN/dx$) and time-of-flight measurements.

At 240~GeV center-of-mass energy and 10.8~ab$^{-1}$ integrated luminosity, the FCC-ee will produce millions of Higgs bosons primarily through Higgs-strahlung (ZH). The clean $e^+e^-$ collision environment enables direct measurement of Higgs branching fractions, contrasting with the ratio measurements at hadron colliders. Advanced flavor-tagging techniques may enable the first observation of Higgs-charm quark coupling.

Regarding invisible Higgs decays, Standard Model coupling measurements suggest that a non-negligible fraction of the total Higgs width could be associated with beyond-the-Standard-Model (BSM) decays~\cite{Patt:2006fw, Argyropoulos:2021sav}. While the Standard Model predicts $H \rightarrow ZZ \rightarrow 4\nu$ with $\mathcal{O}(10^{-3})$ branching fraction, this could increase if Higgs decays into weakly interacting massive particles (WIMPs) \cite{PhysRevLett.103.099905, ELLIS1984453}, potentially explaining dark matter \cite{1937ApJ....86..217Z, 1970ApJ...159..379R, Clowe:2006eq}. Operating at $\sqrt{s} $ 240~GeV and 365~GeV, the FCC-ee could detect branching ratios below one per mille. Current studies examine the ZH process (Z $\to l\bar{l}$ or Z $\to q\bar{q}$ where $l$ is a charged lepton and $q$ is a quark, H $\to$ invisible) using full CLD~\cite{cld} simulation.

Sect.~\ref{sec:idea} details event generation and simulation across detector configurations. Sect.~\ref{sec:tracker} presents extended flavor-tagging performance studies, while Sect.~\ref{sec:ZHfullHad} and~\ref{sec:couplings} discuss implications for Higgs-coupling and self-coupling measurements, respectively. Sect.~\ref{sec:calo} and~\ref{sec:invisible} examine calorimeter variations and their expected impact on Higgs-to-invisible searches, respectively.
 
\section{Generation, simulation and detector variations}\label{sec:idea}
Higgsstrahlung events at a FCC-ee-like Higgs factory ($e^{+}e^{-}\rightarrow ZH$) have been generated through \progname{Whizard}~3.1.2~\cite{whizard,Moretti:2001zz}, at a center-of-mass energy $\sqrt{s} = 240 \text{ GeV}$ and simulated through \textsc{Delphes}~\cite{delphes}, as detailed in Section~\ref{sec:delphes}.

To conduct comprehensive flavor-tagging performance studies, we produced ten million $e^{+}e^{-}\rightarrow Z(\rightarrow\nu\nu)H$ events. These events were evenly distributed among five Higgs decay modes, categorized by jet flavor: bottom ($b$), charm ($c$), strange ($s$), gluon ($g$), and up/down ($q=u,d$). We specifically combined these decay modes with $Z$-boson decays to neutrino pairs ($\nu\nu$) to ensure unambiguous assignment of hadronic products to the Higgs-boson decay.

For the fully hadronic analysis, we utilized additional $e^{+}e^{-}\rightarrow Z(\rightarrow jj)H(\rightarrow jj)$ samples categorized by final state. These include $Z$ decays to $b$, $c$, $s$, and $q$ quarks, and Higgs decays to $b$, $c$, $s$ quarks, gluons ($g$), tau leptons ($\tau$), $Z$ bosons, $Z\gamma$, and $W$ bosons. The primary backgrounds (WW, ZZ, and Zqq) were generated using \progname{Pythia8}~\cite{Bierlich:2022pfr}.

\subsection{IDEA detector concept and \textsc{Delphes} setup}
\label{sec:delphes}

We employed the official \textsc{Delphes} fast simulation framework enhanced with specialized modules for high-precision particle physics simulations: \textsc{TrackCovariance}~\cite{TrackCovariance}, \textsc{TimeOfFlight}~\cite{TimeOfFlight}, and \textsc{ClusterCounting}~\cite{ClusterCounting}, as discussed in Ref.~\cite{FCCFlavTag}. Our baseline configuration adopts the recent IDEA detector prototype~\cite{IDEA1,IDEA2}.

The IDEA conceptual experiment incorporates innovative detector technologies developed in recent years. Its architecture features a Silicon Inner Tracker, encased by a low-mass Drift Chamber and a Silicon wrapper. This inner detector system operates within a 2T magnetic field generated by an ultra-thin superconducting solenoid. A dual-readout calorimeter is positioned beyond the magnet and within its return yoke. For muon detection, the design implements a tracker based on $\mu$-Rwell technology, interleaved within the return yoke material. This same $\mu$-Rwell technology provides an additional pre-shower layer positioned in front of the calorimeter.

To evaluate the impact on flavor-identification performance and Higgs-boson measurements at the FCC-ee, we systematically varied detector properties relative to this baseline in both the inner tracking system and the calorimeter components. These variations are described in detail in subsequent sections.

\subsection{Samples with full simulation}
\label{sec:hinv-fullsim-samples}

For the Higgs-to-invisible analysis, we generated signal samples consisting of $e^{+}e^{-}\rightarrow ZH$ events where the Higgs decays to invisible final states. All samples were produced using \progname{Whizard}~3.1.2~\cite{whizard,Moretti:2001zz} at a center-of-mass energy of $\sqrt{s} = 240 \text{ GeV}$, with hadronization performed by \progname{Pythia}~6~\cite{pythia6}.

For the $Z(H\rightarrow 4\nu)$ signal samples and other $ZH$ backgrounds, we selected specific Higgs boson decay modes using the \verb|unstable| feature of \progname{Whizard}. For $ZZ$ samples, we utilized the ``restriction'' feature of \progname{Whizard} to select the desired processes, with the additional constraint that the generated $Z$ resonance must fall within 5 GeV of the Standard Model value. For $WW$ samples, \progname{Whizard} handled the initial hard scattering, while the subsequent $W$ boson decays were processed by \progname{Pythia}~6.

Cross sections and branching ratios for the $ZH$, $WW$, $ZZ$, and $Z\rightarrow\tau\tau$ samples were taken directly from \progname{Whizard} calculations.

These samples underwent full detector simulation using the standard CLD detector concept~\cite{cld}, implemented through \progname{Geant4}~\cite{geant4} and \progname{DD4hep}~\cite{dd4hep}. Event reconstruction was performed using the \verb|CLDReconstruction.py| and \verb|PandoraSettingsCLD|~\cite{pandorapfa} configurations from the \verb|CLDConfig|~\cite{CLDConfig} repository. All simulation and reconstruction processes utilized the 2023-11-23 release of \progname{Key4Hep}~\cite{key4hep}.

\section{Variations in tracker-detector properties and layout}
\label{sec:tracker}

Jet flavor identification algorithms are essential for interpreting complex event signatures and enabling precise measurements at the FCC-ee. While enhanced tracking resolution improves identification accuracy, it often requires additional detector material (e.g., extra silicon layers), which can reduce the rejection efficiency of unwanted flavors. Understanding and optimizing this trade-off is critical for future collider detector designs.

\subsection{Methodology for performance assessment}

\begin{figure}[hbtp]
    \begin{center}
        \includegraphics[width=0.82\columnwidth]{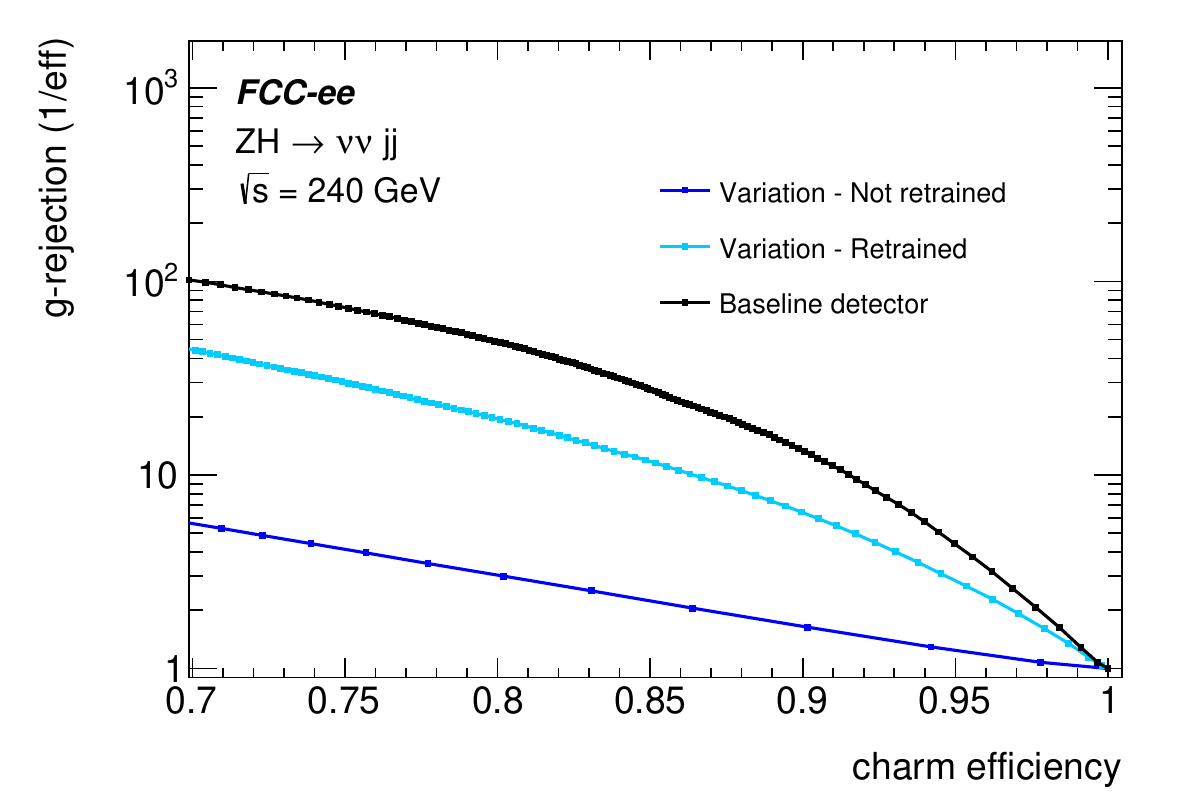}
        \caption{Rejection of gluon-initiated jets as a function of the charm jet-tagging efficiency for the baseline IDEA detector (black), an alternative IDEA tracker without retraining (blue), and the same alternative tracker with retraining (light blue) of the ParticleNet algorithm. Training the algorithm on corresponding detector-configuration features is evidently necessary to properly assess performance potential.}
        \label{fig:retraining}
    \end{center}
\end{figure}

Each detector variation necessitates retraining of jet flavor identification algorithms to ensure fair and meaningful performance assessment. Figure~\ref{fig:retraining} illustrates the Receiver Operating Characteristic (ROC) curves for both the baseline detector and a detector variation, with and without algorithm retraining on samples accounting for such variations.

An algorithm trained on samples simulated with one detector configuration but evaluated on samples from an alternative configuration can significantly misrepresent the magnitude of performance differences. Performance degradation caused by detector limitations can be largely mitigated by retraining the algorithm on appropriately simulated samples which include those limitations. Thus, all tagger-performance studies presented herein utilize algorithms trained and evaluated on consistent input samples.

\subsection{Tracker single-point resolution}

The baseline IDEA detector simulations assume a nominal single-point resolution of $3\,\mu\text{m}$, achievable with a $25\times25\,\mu\text{m}^2$ pitch vertex detector~\cite{ALICEITS3}. We investigated the impact of varying this resolution by $\pm65\%$, a direct consequence of hit resolution modifications, and found it significantly affects jet flavor determination.

\begin{figure}[hbtp]
    \includegraphics[width=0.49\columnwidth]{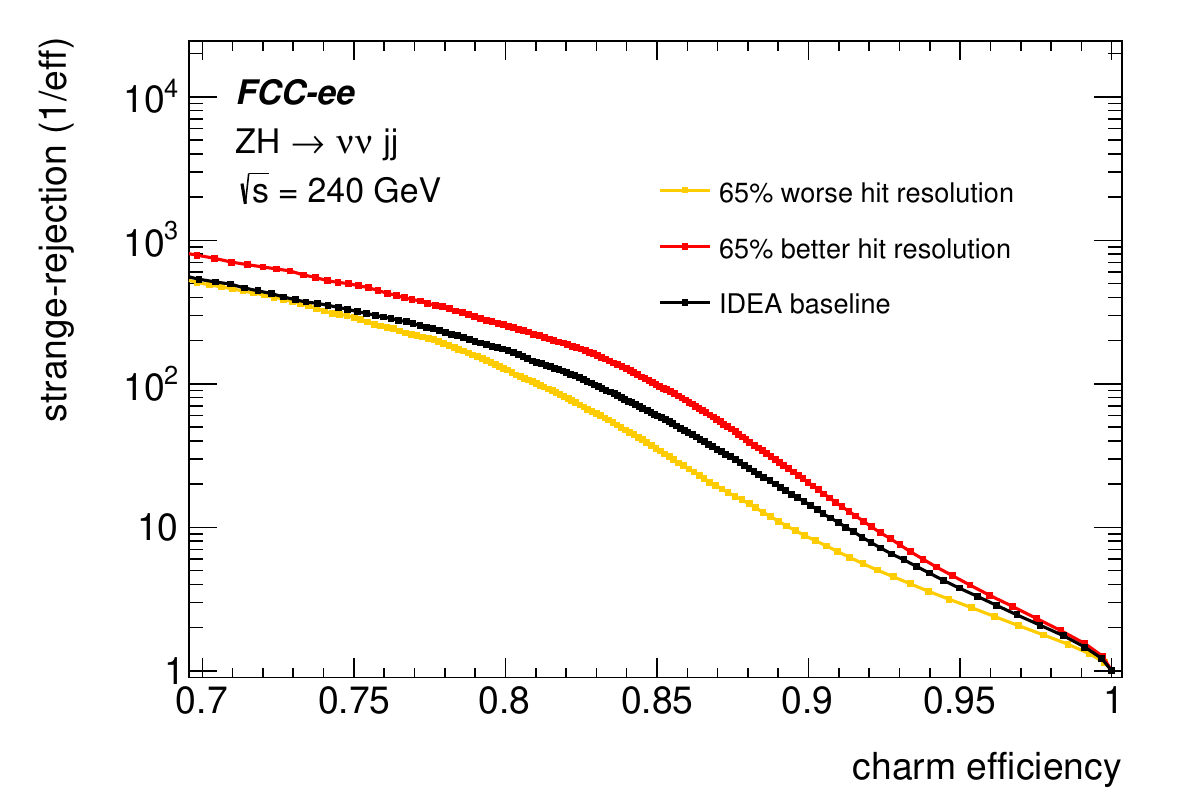}
    \includegraphics[width=0.49\columnwidth]{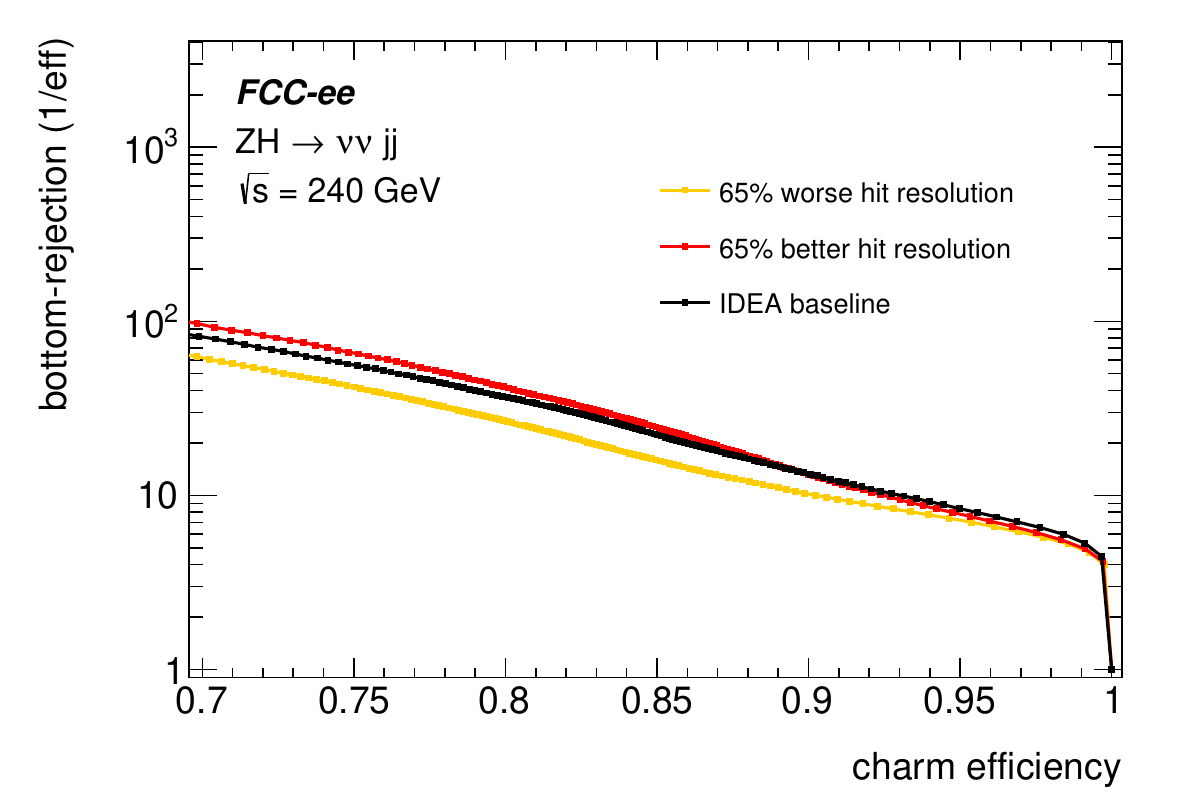}
    \caption{Rejection of strange-initiated jets (left) and bottom-initiated jets (right) as a function of charm jet tagging efficiency for different hit resolutions.}
    \label{fig:sc_bc_pixresol}
\end{figure}

Figure~\ref{fig:sc_bc_pixresol} shows that $s$-jet rejection is highly sensitive to single-point resolution. Effects exhibit symmetry in the higher $c$-efficiency range, with approximately a factor of 2 improvement or degradation in $s$-rejection capability depending on resolution changes.

Conversely, the rejection of $b$-jets for $c$-tagging shows asymmetric effects, suggesting that asymptotic momentum resolution may have limited relevance at the FCC-ee due to the relatively low transverse momenta ($p_{\text{T}} \leq 50 \text{ GeV}$) of most tracks in jets. As Figure~\ref{fig:sc_bc_pixresol} demonstrates, improved tracker resolution provides minimal benefit for $b$-jet rejection, which is crucial for $H\to cc$ analyses. Similar effects are observed when rejecting jets from other flavors while tagging $c$-jets. For $b$- and $s$-tagging, the impacts are considerably smaller, with relative changes in rejection capabilities for other flavors remaining within $10$-$20\%$.

\subsection{Silicon and beam-pipe material budget}

While improved tracking resolution can enhance identification accuracy, it frequently necessitates additional detector material. Here we examine how varying the vertex detector material content affects performance, along with the interplay between beam-pipe and vertex-detector material budgets.

Assuming a standard radiation length for silicon of 0.0937 m, we studied the effect of simultaneously varying the radiation length for all silicon layers, equivalent to modifying the material content by $\pm50\%$. For $c$-tagging, we observed an asymmetric impact: negative effects from additional material were five times larger than benefits from reduced material. The maximum observed change in strange-jet rejection capability while tagging $c$-jets was approximately $30\%$. Such variations had only minor effects on $b$-jet and strange-jet tagging performance.

These results indicate that while a lighter vertex detector offers minimal performance gains, additional or denser material can substantially degrade performance. For jets with momentum above 70 GeV, as expected, the impact of multiple Coulomb scattering diminishes significantly. Moreover, with substantial increases in beam-pipe material budget, changes in the vertex-detector first layer material budget become secondary concerns.

\subsection{Removal of intermediate silicon layers}

Previous studies documented in Ref.~\cite{FCCFlavTag} demonstrate that an additional innermost silicon layer plays a key role in $c$-tagging, providing minor improvements in $b$-tagging and enhanced background rejection in $c$-tagging. In the current IDEA baseline detector concept, the beam pipe has been reduced to 1~cm diameter, with the four innermost silicon layers positioned at 1.2, 2, 3.15, and 15~cm from the beam axis. We evaluated the impact on jet flavor tagging of removing either the second or fourth innermost silicon layer.

Our studies revealed measurable effects on $c$-tagging performance, with removal of the second innermost silicon layer producing a more significant impact than removal of the fourth layer. Relative changes in background jet rejection remained within $15\%$, with similar patterns observed for $s$-, $q$-, and $g$-jet rejection while tagging $c$-jets. While $b$-tagging appears more robust against these changes, the effects may amplify when analyzing jets produced at higher momenta, such as in differential measurements.

\subsection{Drift chamber particle identification capabilities}
\label{sec:driftCh}

To characterize flavor-tagging performance enhancements from particle identification (PID) capabilities, we investigated several variations in PID information provided to the algorithm training: baseline PID, complete absence of PID, absence of cluster-counting information, absence of time-of-flight (TOF) measurement, absence of both TOF and cluster-counting information, and baseline PID with improved timing resolution.

\begin{figure}
    \includegraphics[width=0.49\linewidth]{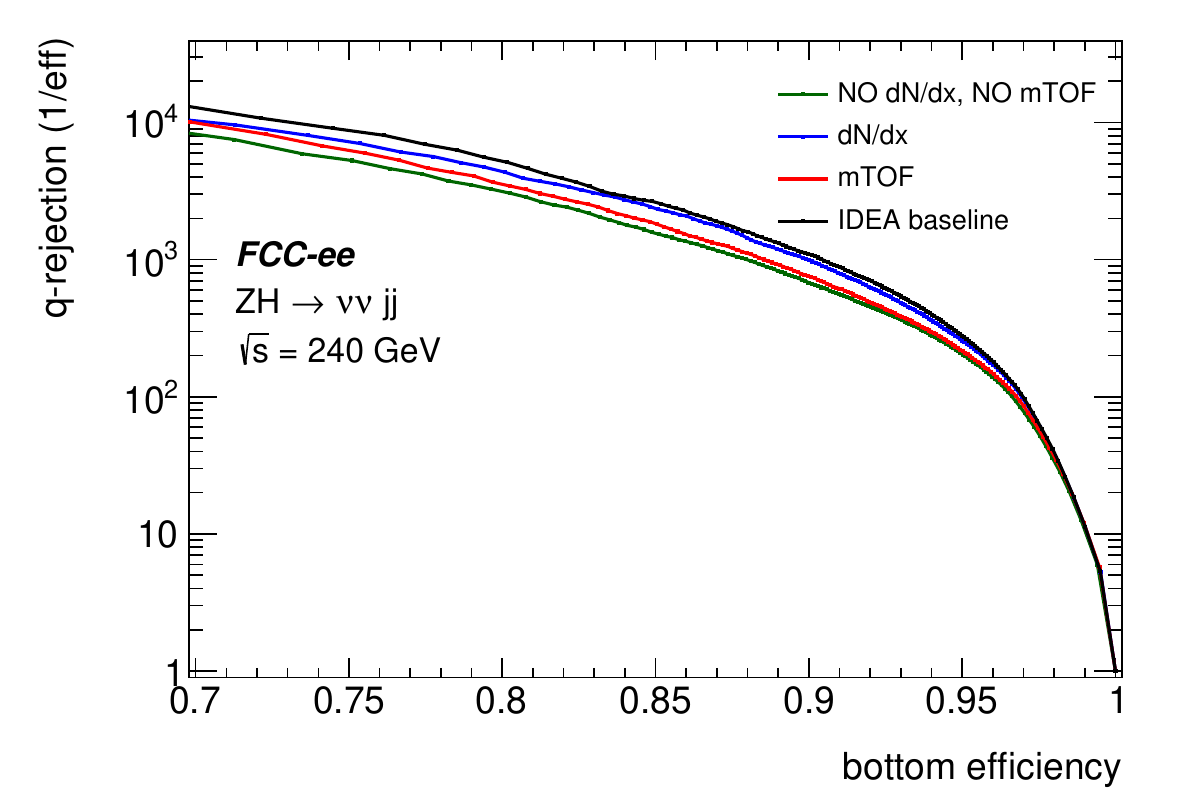}
    \includegraphics[width=0.49\linewidth]{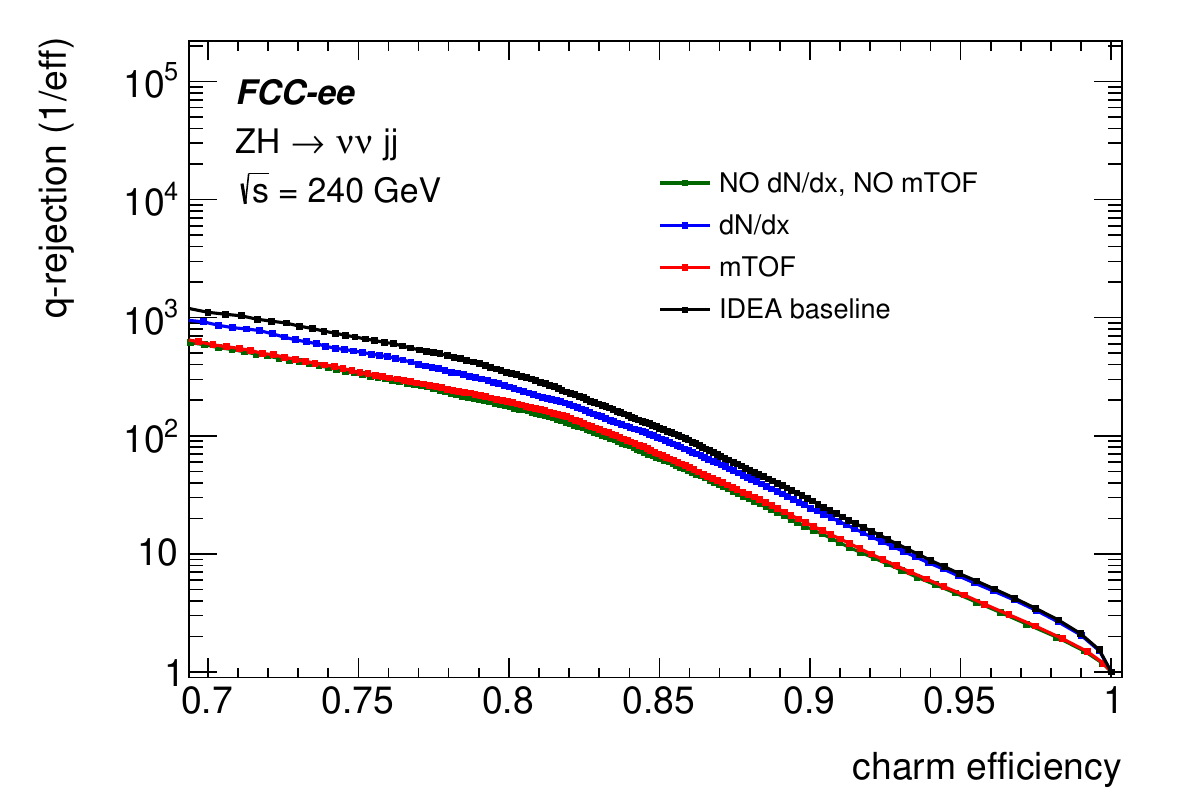}
    \centering
    \includegraphics[width=0.49\linewidth]{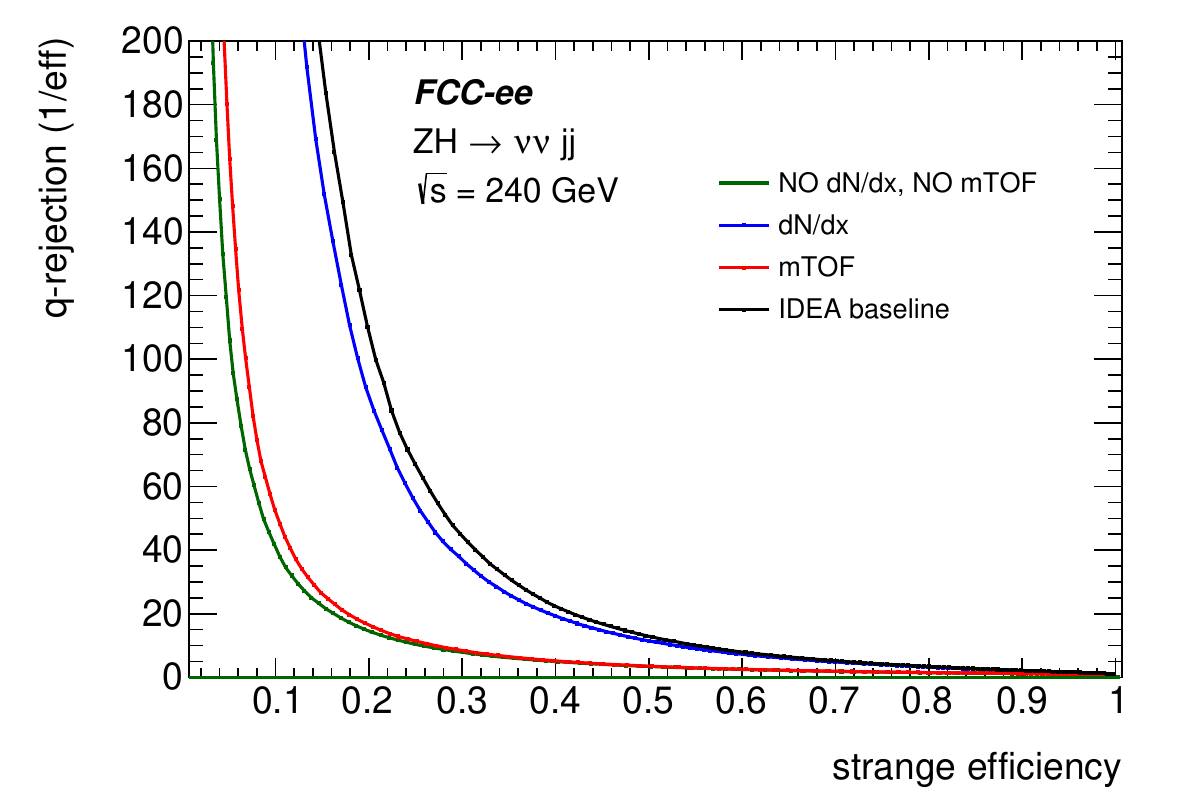}
    \caption{Rejection of light-initiated jets as a function of jet tagging efficiency for bottom (upper left), charm (upper right), and strange (lower) jets. Results are shown for the baseline IDEA vertex detector (black), without cluster counting (red), without TOF information (blue), and without both PID components (green).}
    \label{fig:qs_PID}
\end{figure}

Figure~\ref{fig:qs_PID} shows that timing measurements contribute minimally to strange-jet tagging in this kinematic regime. As discussed in Ref.~\cite{FCCFlavTag}, TOF primarily contributes to kaon-pion separation at momenta around 1 GeV. In contrast, TOF measurements demonstrate noticeable effects on charm-jet tagging and significant impacts on bottom-jet tagging, both independently and combined with cluster-counting information.

\begin{figure}
    \includegraphics[width=0.49\linewidth]{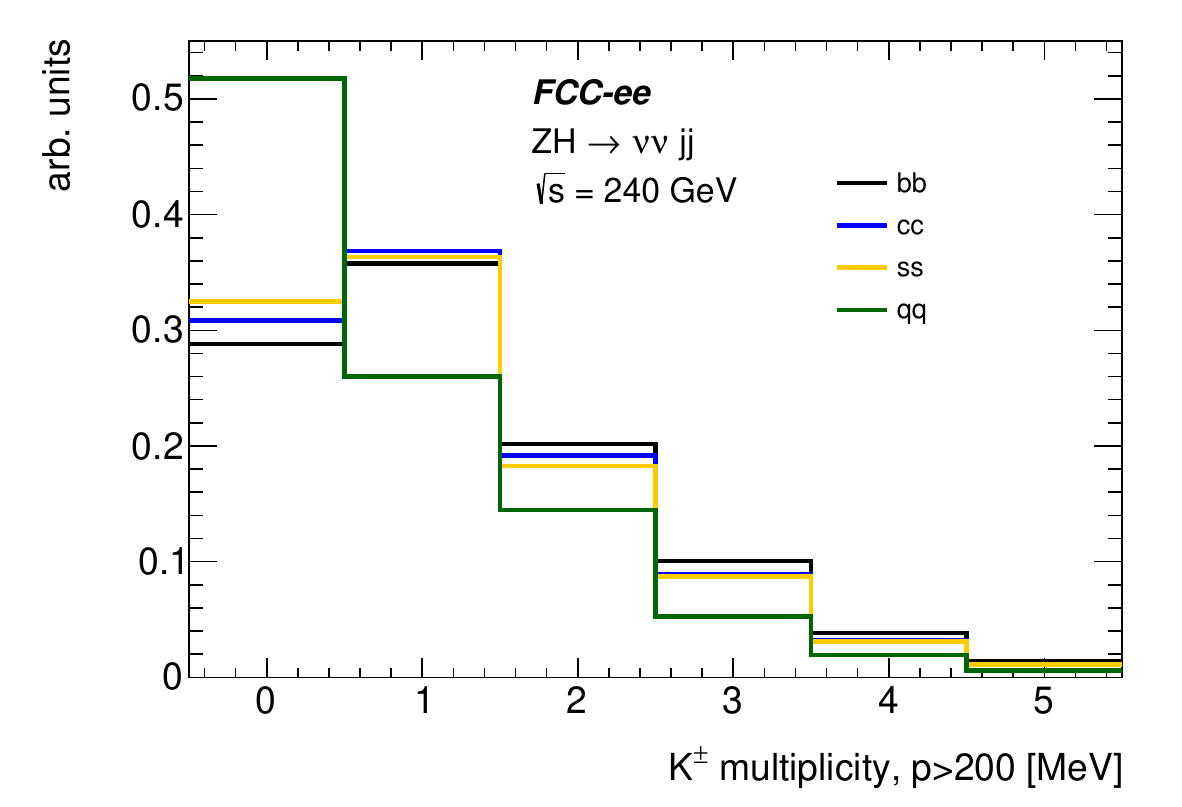}
    \includegraphics[width=0.49\linewidth]{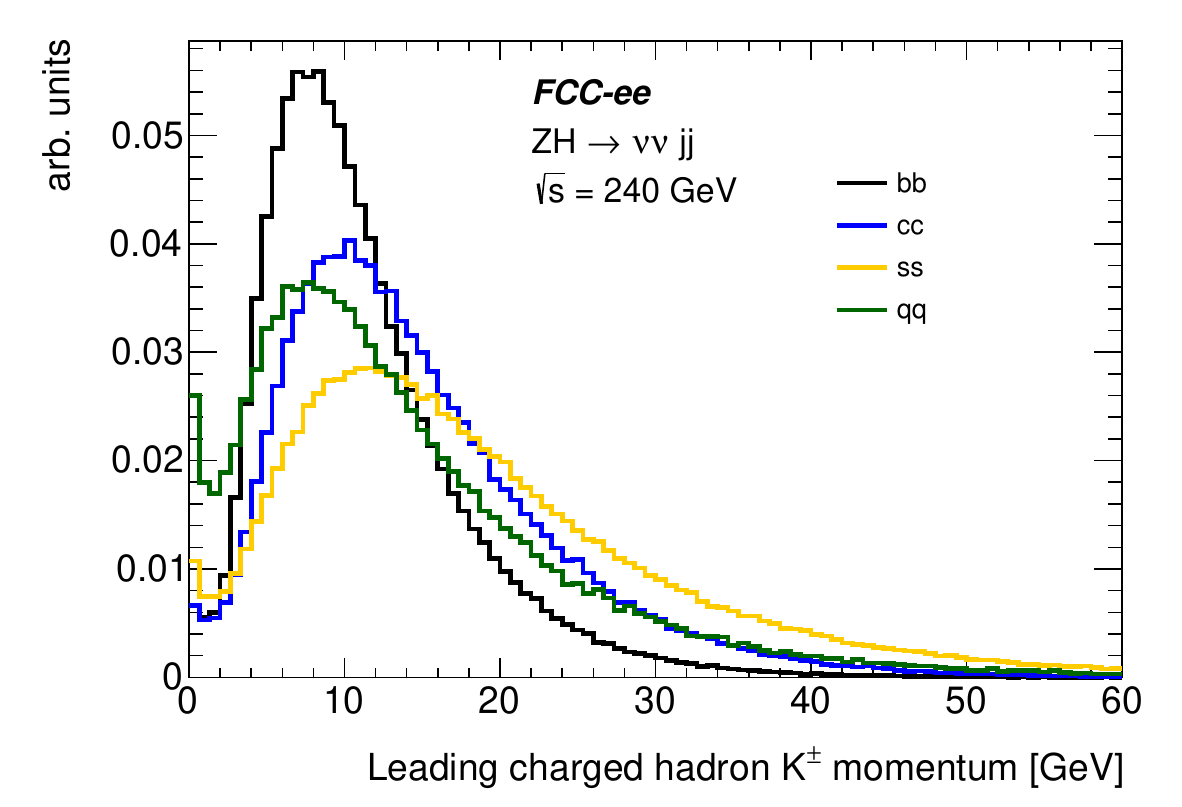}
    \caption{Multiplicity of kaons with minimum momentum of 200 MeV (left) and momentum distribution of the leading kaon (right) within bottom (black), charm (blue), strange (yellow), and light (green) jets.}
    \label{fig:nK_leadKp}
\end{figure}

To establish an upper performance limit achievable with PID information, we implemented truth-matching of particle-flow candidates to Monte Carlo (MC) truth particles. Figure~\ref{fig:nK_leadKp} presents the multiplicity of truth-matched kaons and the momentum distribution of the leading kaon within Higgs-decay jets of different flavors.

\begin{figure}
    \includegraphics[width=0.49\linewidth]{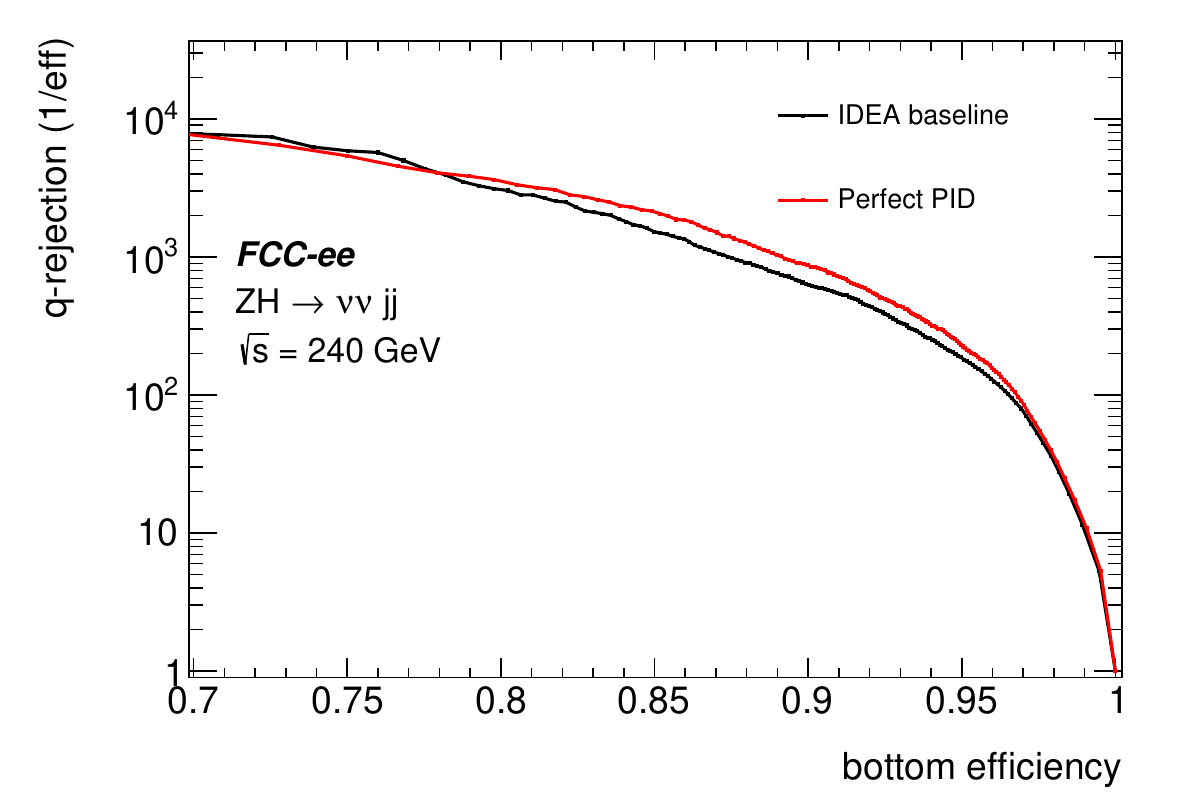}
    \includegraphics[width=0.49\linewidth]{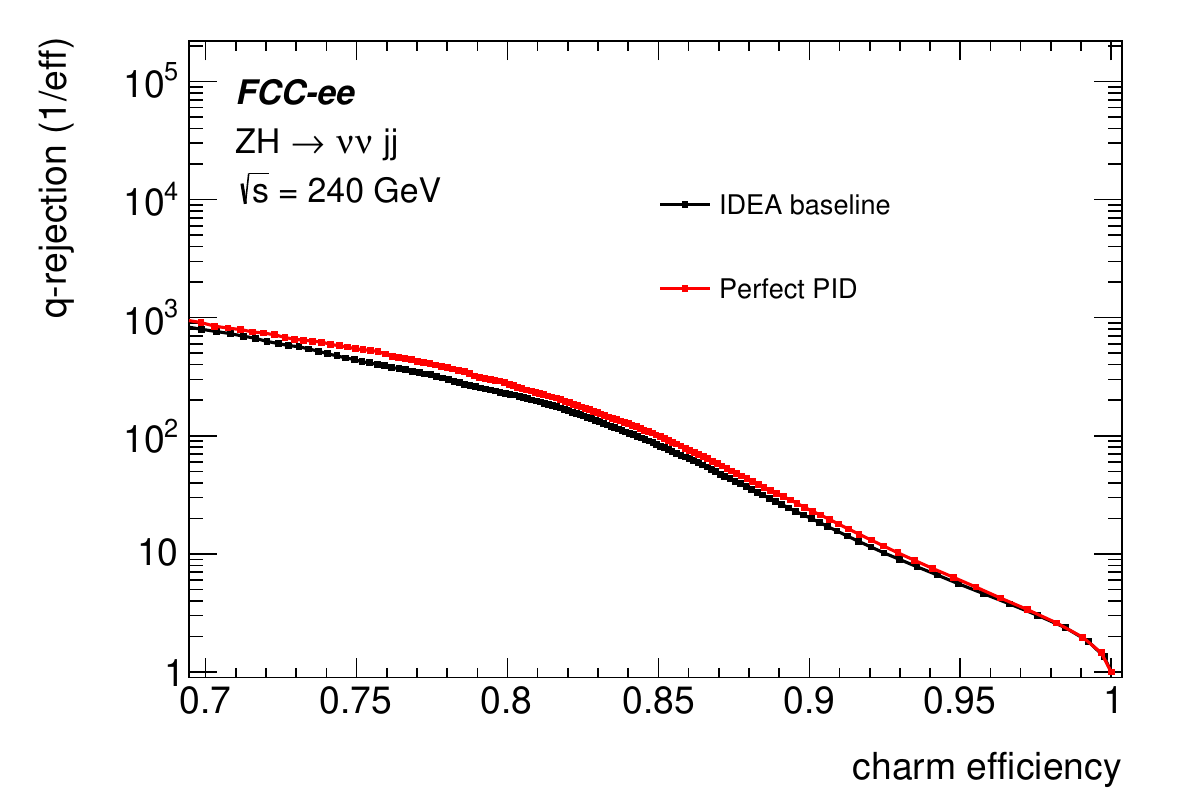}
    \centering
    \includegraphics[width=0.49\linewidth]{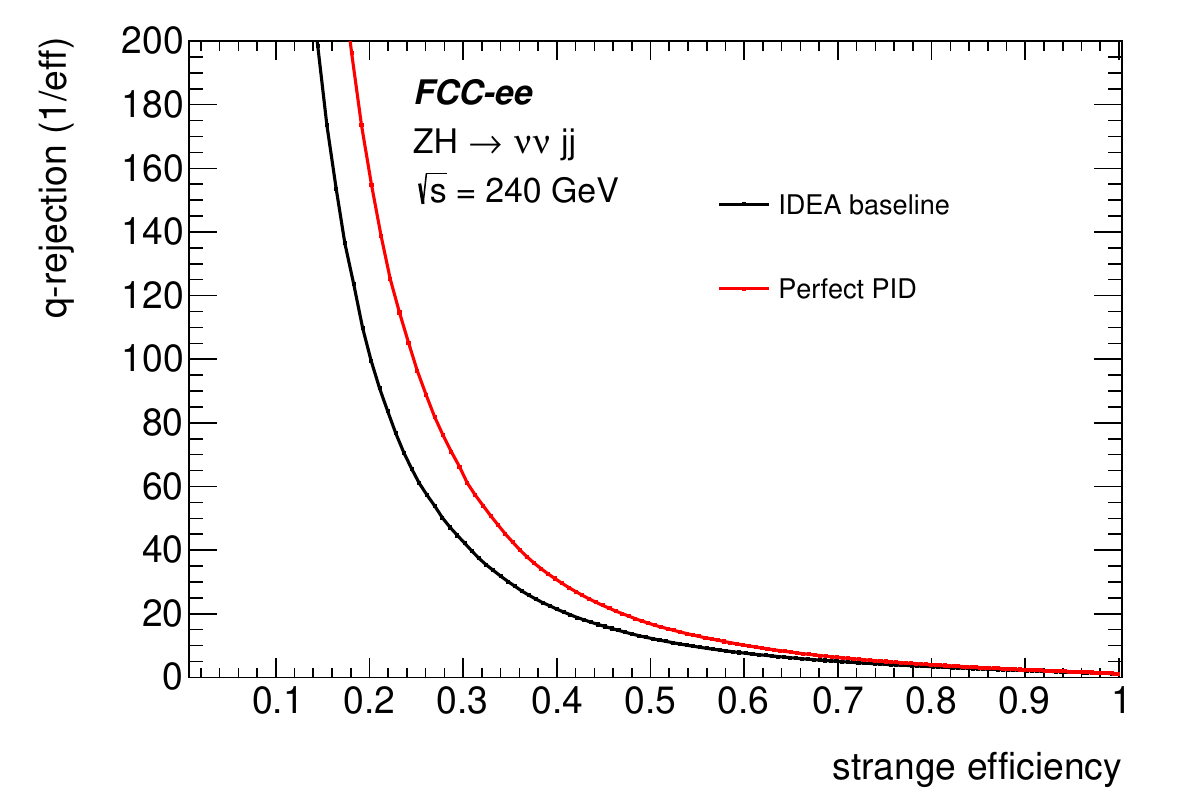}
    \caption{Light-jet rejection performance for bottom tagging (upper left), charm tagging (upper right), and strange tagging (bottom). Results compare the baseline training (black) with training incorporating truth-MC information for charged-hadron and charged-lepton particle-flow candidates (red).}
    \label{fig:truthPID_ROCs}
\end{figure}

We incorporated truth-MC information identifying charged-hadron and charged-lepton particle-flow candidates in a dedicated training. Figure~\ref{fig:truthPID_ROCs} shows the resulting performance improvements. Bottom (charm) jets exhibit a larger fraction of leading charged kaons around 1 GeV than charm (strange) jets, correlating with the hierarchy of performance effects observed in tagger performance.

\section{Variations in properties of calorimeter detectors}
\label{sec:calo}

The nominal energy resolution for the IDEA prototype calorimeters has been assumed from Ref.~\cite{2008.00338}.
In the electromagnetic crystal calorimeter:
\begin{equation}
    \sqrt{a^2E^2 + b^2E + c^2}
\end{equation}
with $a=0.005$, $b=0.03$ and $c=0.002$.

In the hadronic dual-readout calorimeter:
\begin{equation}
    \sqrt{d^2E^2 + e^2E + f^2}
\end{equation}
with $d=0.01$, $e=0.3$ and $f=0.05$.
The nominal granularity of calorimeters is simulated by means of a cell size of about $6\times6~cm^2$.

Energy-resolution parameters ($a,b,c,d,e,f$) are scaled up and down at the same time by a relative $30\%$ to mimic potential degradation and improvement in the calorimeter performance. The nominal cell size is varied by a factor 2 in order to estimate the potential of a more or less granular calorimetry system.
 
\section{Statistical methodology and limits}
\label{sec:stats}

To determine upper limits on the branching ratio ($\text{BR}$) of the process under study, we employ the \texttt{xRooFit} toolkit~\cite{xRooFit}, considering only statistical uncertainties. We construct a likelihood function using signal and background models from simulation, expressing the total probability density function (PDF) as:
\begin{equation}
P(x \mid \mu) = \mu \cdot S(x) + (1 - \mu) \cdot B(x),
\end{equation}
where $x$ represents the observable variable (such as recoil mass), $S(x)$ and $B(x)$ are the signal and background components, and $\mu$ is the signal strength parameter related to the branching ratio by:
\begin{equation}
\mu = \frac{\sigma_{\text{obs}} \cdot \text{BR}}{\sigma_{\text{SM}}},
\end{equation}
with $\sigma_{\text{obs}}$ being the observed production cross-section and $\sigma_{\text{SM}}$ the Standard Model prediction.

We generate pseudo-datasets by sampling the expected signal-plus-background or background-only distributions, incorporating statistical fluctuations consistent with expected event yields. The profile likelihood ratio served as our test statistic:
\begin{equation}
\lambda(\mu) = \frac{\mathcal{L}(\mu)}{\mathcal{L}(\hat{\mu})},
\end{equation}
where $\mathcal{L}(\mu)$ is the likelihood given the signal strength $\mu$, and $\hat{\mu}$ maximizes the likelihood. We evaluated $-2\ln\lambda(\mu)$ to compare signal-plus-background and background-only hypotheses, determining its distribution under the background-only hypothesis using asymptotic approximations. 

For each detector variation described in previous sections, we performed separate likelihood scans to determine the impact on sensitivity. These scans were conducted by fixing all nuisance parameters to their best-fit values and varying only the signal strength parameter $\mu$. For analyses with multiple channels, we constructed a joint likelihood incorporating all relevant signals and backgrounds, accounting for their respective efficiencies and resolutions. The 95\% confidence level (CL) upper limit on the parameter of interest corresponds to the $\mu$ value where the cumulative probability exceeds the critical threshold. This methodology is applied consistently across the analyses described in the following sections.

\section{ZH analysis in the fully hadronic final state}
\label{sec:ZHfullHad}
We reran the ZH fully hadronic analysis for each retrained flavor tagger, using the expected 68\% confidence level (CL) uncertainty on the measured Higgs couplings to $b$-, $c$-, $s$-quarks, and gluons as metrics to determine the impact of flavor tagger performance. The analysis methodology follows that described in Ref.~\cite{GeorgesPaper1} and is summarized below.

\subsection{Analysis methodology}

The exclusive Durham $k_t$-algorithm is employed for jet clustering, requiring exactly four jets per event corresponding to the hadronic decays of both $Z$ and $H$ bosons. To suppress backgrounds, we apply cuts on angular separation between jets: $15000 < d_{12} < 58000$, $400 < d_{23} < 18000$, and $100 < d_{34} < 6000$, where $d_{ij} = 2\min(E_i^2,E_j^2)(1-\cos\theta_{ij})$ between the $i^{\text{th}}$ and $j^{\text{th}}$ jets. Additionally, we require visible energy and mass greater than 150~GeV, visible $\theta$ within $[0.15, 3.0]$, and at most two leptons with the leading lepton momentum below 20~GeV. Assuming well-measured jet directions and known initial energy in a four-jet system, jet energies are corrected following energy and momentum conservation principles.

Flavor scores from the tagger are used to pair jets and identify $Z$ or $H$ candidates based on dijet masses. Each jet has flavor scores $B_i$, $C_i$, $S_i$, $Q_i$, and $G_i$ corresponding to the probabilities of the jet being initiated by $b$-, $c$-, $s$-, $u/d$-quarks, or gluons, respectively. We identify the maximum flavor score for each jet and apply different pairing strategies depending on the pattern of maximum scores:

\begin{description}
\item[Case 1 - All jets with the same maximum flavor score:]
To determine which dijet pair forms the $H/Z$ candidate, we consider all possible jet pairings and compute $\chi^2_{\text{comb}} = \chi^2_Z + \chi^2_{\text{Higgs}}$, where $\chi^2_Z = (m_{Z,\text{reco}} - m_{Z,\text{true}})^2$ and $\chi^2_{\text{Higgs}} = (m_{H,\text{reco}} - m_{H,\text{true}})^2$. The jet pairing with minimum $\chi^2_{\text{comb}}$ is selected as correct, where $m_{Z/H,\text{reco}}$ are the reconstructed $H/Z$ dijet masses.

\item[Case 2 - Pairs with different maximum flavor scores:]
If combining jets with the same maximum flavor scores yields exactly two pairs with different flavors, the $Z$ candidate is chosen as the pair with minimum $\chi^2_Z = (m_{lk} - m_{Z,\text{true}})^2$, and the remaining pair is assigned as the $H$ candidate.

\item[Case 3 - Two jets with the same maximum flavor score:]
Two jets with maximum scores from the same flavor are paired, while the flavor of the second pair is determined by examining same-flavor score sums. The flavor of the second pair is chosen as the one that maximizes the same-flavor sum, $\max(\sum_{i,j}B, \sum_{i,j}C, \ldots)$. The $H/Z$ candidates are then determined as in Case 1 if both pairs have the same flavor, or as in Case 2 if they have different flavors.

\item[Case 4 - Three jets with the same maximum flavor score:]
Similar to Case 3, the first pair is formed from the jets whose same-flavor sum of scores is greatest. The flavor of the second pair is determined as in Case 3, and the assignment of $H/Z$ candidates follows the approach outlined in Case 3.

\item[Case 5 - Four different maximum flavor scores:]
If each jet has a maximum flavor score from a different flavor, the event is excluded from analysis.
\end{description}

To reject $WW$ and $ZZ$ backgrounds, we apply additional cuts based on the $H$ and $Z$ candidate masses. The final fit is performed in the $H$ and $Z$ mass plane, where $50~\text{GeV} < m_{Z,\text{reco}} < m_{H,\text{true}}$ and $m_{H,\text{reco}} > m_{Z,\text{true}}$. We classify events into 60 categories based on the flavor of the $H$ and $Z$ decay products and the sum of flavor scores for the pair forming the Higgs candidate (purity categories).

\subsection{Impact of flavor tagger performance}
\label{sec:impactofTaggerPerf}

While Section~\ref{sec:tracker} demonstrated that tagger performance may be overestimated when evaluated on simulations different from those used in training, this effect has negligible impact at the analysis level. We tested this by applying both the ParticleNet tagger trained on the baseline IDEA simulation and the tagger trained on a simulation with 65\% worse single-hit resolution to $Zcc$-$Hbb$, $Zcc$-$Hss$, and $Zcc$-$Hgg$ signal samples generated with both detector geometries. The Higgs decay tagging efficiency showed no dependence on which tagger was applied, likely due to the robustness of our flavor tagging strategy. The effects observed in Section~\ref{sec:tracker} are most prominent at lower tagging efficiencies, whereas our analysis ensures high efficiency by using the sum of dijet flavor scores.

\begin{figure}
    \begin{center}
    \subfloat[$\mu_{Hbb}$]{\includegraphics[width=0.45\textwidth]{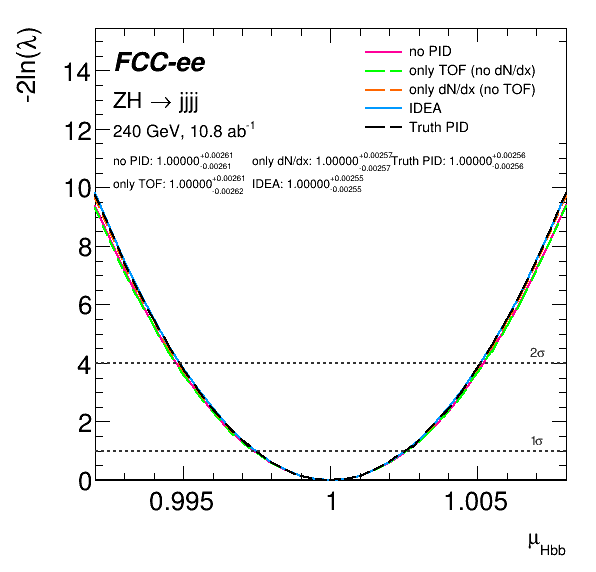}}
    \subfloat[$\mu_{Hcc}$]{\includegraphics[width=0.45\textwidth]{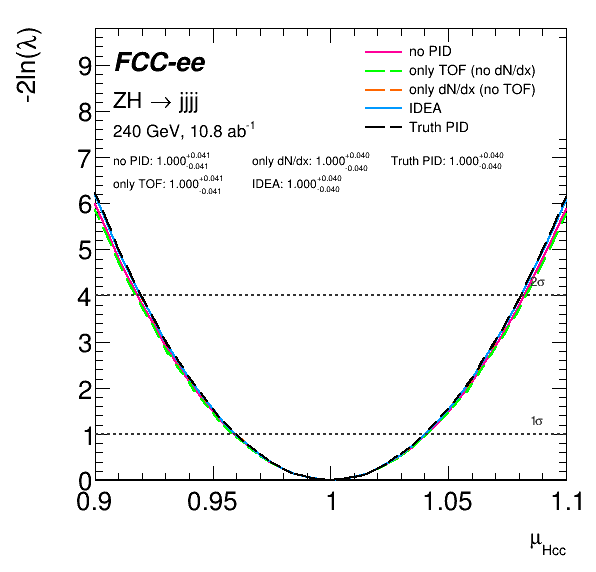}}\\
    \subfloat[$\mu_{Hss}$\label{fig:HssScan}]{\includegraphics[width=0.45\textwidth]{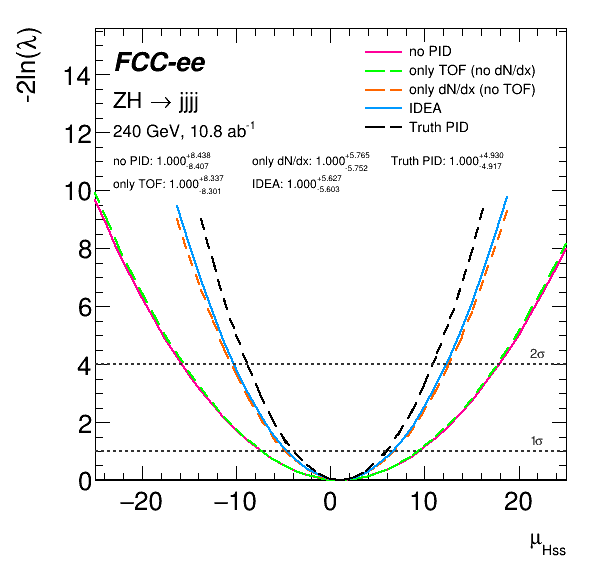}}
    \subfloat[$\mu_{Hgg}$]{\includegraphics[width=0.45\textwidth]{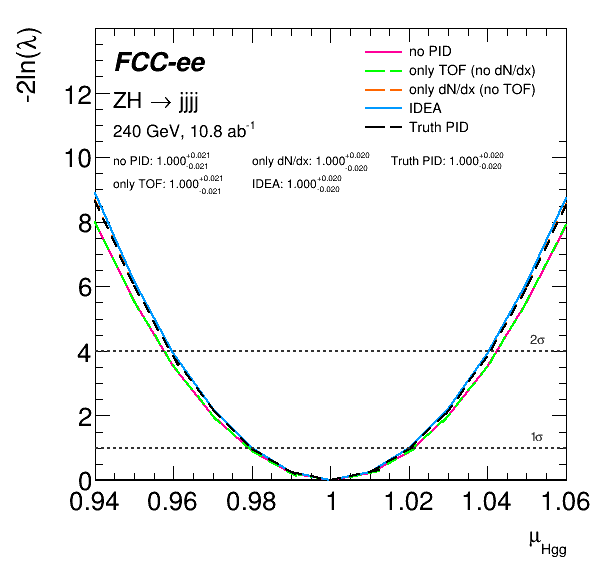}}
    \end{center}
    \caption{Impact of flavor tagger training variations on negative log-likelihood (NLL) scans for Higgs coupling measurements. The baseline training (IDEA, blue) and training without time-of-flight information and only the dN/dx information (only dN/dx, orange) show similar performance when applied to the ZH fully hadronic analysis. However, removing cluster counting information and keeping the TOF information (no TOF, green) significantly degrades the precision of $\mu_{Hss}$ measurement similarly as having no PID information at all (no PID, pink). Including truth information in the training of the flavour tagger (Truth PID, black) modestly improves precision for $\mu_{Hcc}$ and $\mu_{Hss}$ measurements, with no significant effect on $\mu_{Hbb}$ and $\mu_{Hgg}$.}
    \label{fig:ZHScans}
\end{figure}

Figure~\ref{fig:ZHScans} illustrates the impact of flavor tagger performance on likelihood scans and the resulting 68\% CL uncertainties for Higgs couplings to $b$-, $c$-, $s$-quarks, and gluons. We compare the baseline IDEA detector tagger with variants trained without time-of-flight information (no TOF, only dN/dx), without cluster counting (no dN/dx, only TOF), without both dN/dx and TOF ( no PID), and with truth-level particle identification (Truth PID). Notably, detector variations including removal of the innermost silicon layer, displacement of silicon barrel layers, 50\% heavier or lighter vertex detector, or 65\% better/worse single-hit resolution did not significantly affect Higgs coupling precision, consistent with expectations from the ROC curves as can be seen in Figs.~\ref{fig:sc_bc_pixresol} for example.

The Higgs decay flavor reconstruction remains largely unchanged when using a tagger trained without time-of-flight information. This robustness arises from our flavor tagging technique, which determines $H/Z$ flavor with high efficiency regardless of detector geometry variations, as evidenced by the ROC curves in Fig.~\ref{fig:qs_PID}.

However, removing cluster-counting information from tagger training significantly impacts Higgs coupling measurements. Without cluster counting, the expected uncertainty on $\mu_{H\rightarrow s\bar{s}}$ at 68\% Confidence Level is 1.5 times worse as compared to the baseline uncertainty. The results are as expected from a closer inspection of Figure \ref{fig:qs_PID}, which shows that the light jet rejection as a function of s-tagging efficiency has a notable impact from the inclusion of cluster-counting information, as compared to missing TOF information. The smaller impact of removing cluster-counting information on the measured $\mu_{Hbb}$ and $\mu_{Hcc}$, as compared to $\mu_{Hss}$, is because the light jet rejection is significantly lower at high s-efficiency in contrast to the behavior observed at high b- and c-efficiencies.

\subsection{Comparing clustering algorithms in fully hadronic ZH events}

\label{sec:Clusteringalgs}

The exclusive Durham-$k_t$ jet clustering algorithm configured in $n$-jets mode has been the standard approach for jet reconstruction throughout the FCC feasibility studies. However, we observed jet pair misassignment and soft particle misclustering in $e^{+}e^{-}\rightarrow ZH$ Higgs production when using this algorithm. To address these limitations, we implemented and evaluated the anti-$k_t$ clustering algorithm as an alternative for reconstructing Higgs mass in the fully hadronic 4-jet final state.

We selected the inclusive anti-$k_t$ algorithm as a promising alternative to Durham-$k_t$ because of its infrared and collinear safety properties, which ensure reliable performance when processing soft or closely spaced particles. For our comparative analysis, we used $ZH$ events where the Higgs boson decays to a $b\bar{b}$ pair and the $Z$ boson decays to a $c\bar{c}$ pair, providing a clean fully hadronic 4-jet final state.

For events with more than four jets, we applied an energy recovery algorithm following the procedure outlined in Ref.~\cite{JetAlgorithms}. This procedure first selects the four highest-energy jets and then merges each remaining soft jet with the hard jet closest in angular separation. After merging, we applied a 10 GeV minimum energy threshold to all jets. We then processed the data using the anti-$k_t$ clustering algorithm with jet radius parameter $R$ varying from 0.4 to 1.0.

\begin{figure}[hbtp]
    \begin{center}
        \includegraphics[width=0.75\columnwidth]{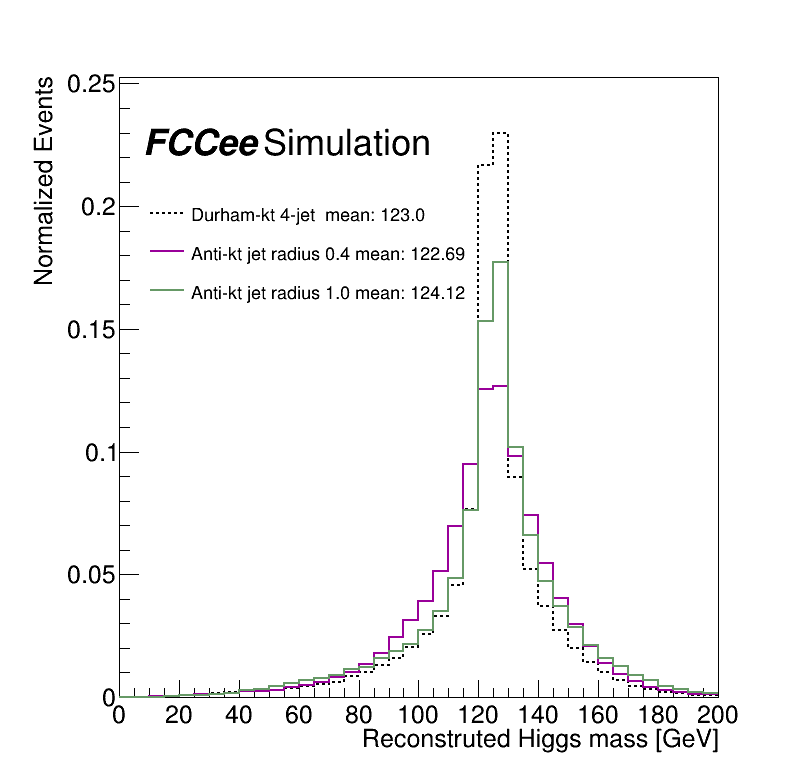}
        \caption{Comparison of reconstructed Higgs boson mass distributions using the Durham-$k_t$ clustering algorithm configured for exactly 4 jets and the anti-$k_t$ clustering algorithm with jet radius parameters $R=0.4$ and $R=1.0$. The anti-$k_t$ algorithm with $R=1.0$ yields a distribution centered closest to the true Higgs mass.}
        \label{fig:jetclustercompare} 
    \end{center}
\end{figure}

Figure~\ref{fig:jetclustercompare} presents the reconstructed Higgs boson mass distributions using different clustering approaches. The anti-$k_t$ algorithm with radius parameter $R = 1.0$ produced the best mass resolution, yielding a mean Higgs invariant mass of 124.1 GeV, which is closer to the true Higgs mass than the 123.0 GeV mean value obtained with the Durham-$k_t$ algorithm. The optimal performance of the $R = 1.0$ configuration is likely related to the event topology in $e^{+}e^{-}$ collisions at $\sqrt{s} = 240$ GeV, where the relatively low boost of the Higgs boson results in decay products with wider angular separation compared to high-energy hadron collider environments. \textbf{Further studies are underway to provide a more detailed understanding of this radius dependence.}

\section{ZH analysis in the \texorpdfstring{$Z\to\nu\bar{\nu},H(jj)$}{Z->nu nubar, H(jj)} hadronic final state}
\label{sec:nunujj}
\begin{figure}[hbtp]
    \includegraphics[width=0.49\columnwidth]{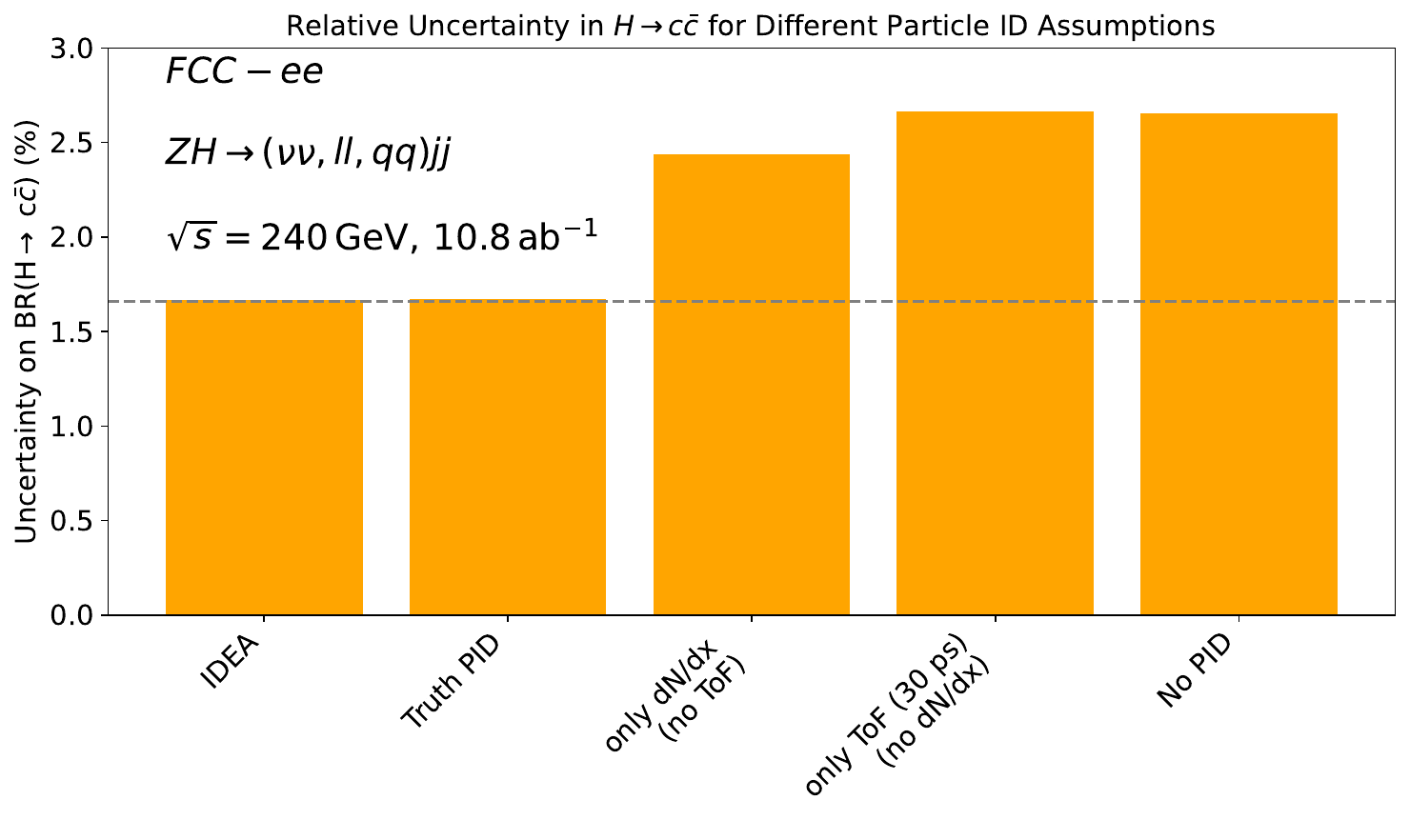}
    \includegraphics[width=0.49\columnwidth]{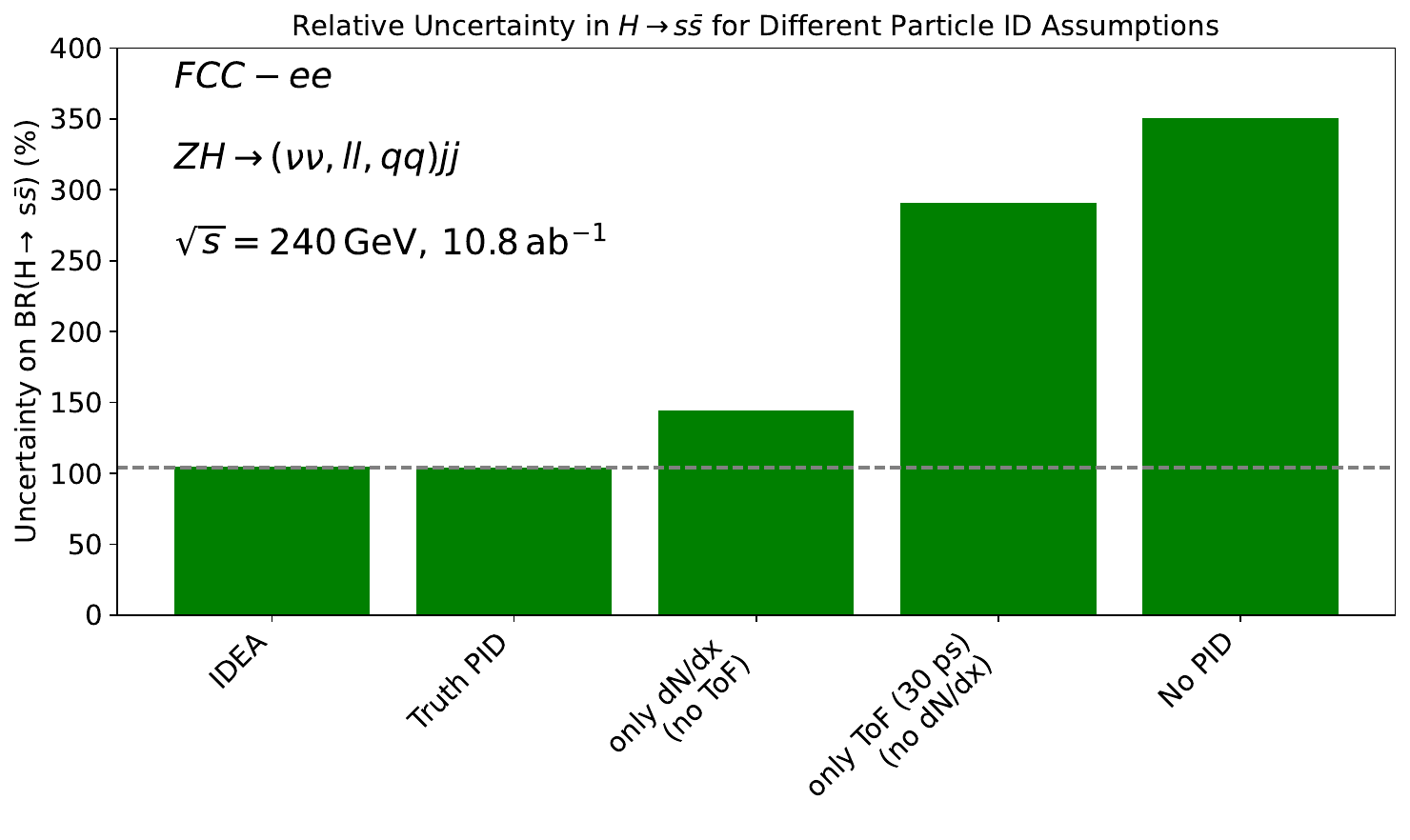}
    \caption{Relative uncertainty in the $H\to c\bar{c}$(left) and $H\to s\bar{s}$ (right) branching fractions, as expected for several
assumptions on particle identification performance.}
    \label{fig:nunujjpidvariation}
\end{figure}

The anticipated precision on the Higgs boson branching fractions to $c\bar{c}$ and $s\bar{s}$ for several particle identification capabilities has been also explored using the $Z(\nu \bar{\nu})H(jj)$ topology. The analysis strategy leverages the flavor tagging capabilities of the ParticleTransformer tagger in a relatively clean reconstruction environment where the final state consists of only two jets of the same flavor and missing mass from the neutrinos.  The analysis is described in detail in \cite{GeorgesPaper1}. The results are shown in Fig.~\ref{fig:nunujjpidvariation}. 
As anticipated, efficient $K/\pi$ separation is crucial for tagging strange quarks effectively. Furthermore, cluster counting (d$N/$d$x$)provides significant separation power, resulting in higher precision in the relative uncertainty. If only Time of Flight (ToF) measurements are assumed, the precision of the $H\to s\bar{s}$  branching fraction measurement severely degrades to values exceeding 300\%. The particle identification also contributes to the  $H\to c\bar{c}$  branching fraction measurement through the decay products of $D$ mesons.

\begin{figure}[hbtp]
\centering
    \includegraphics[width=0.49\columnwidth]{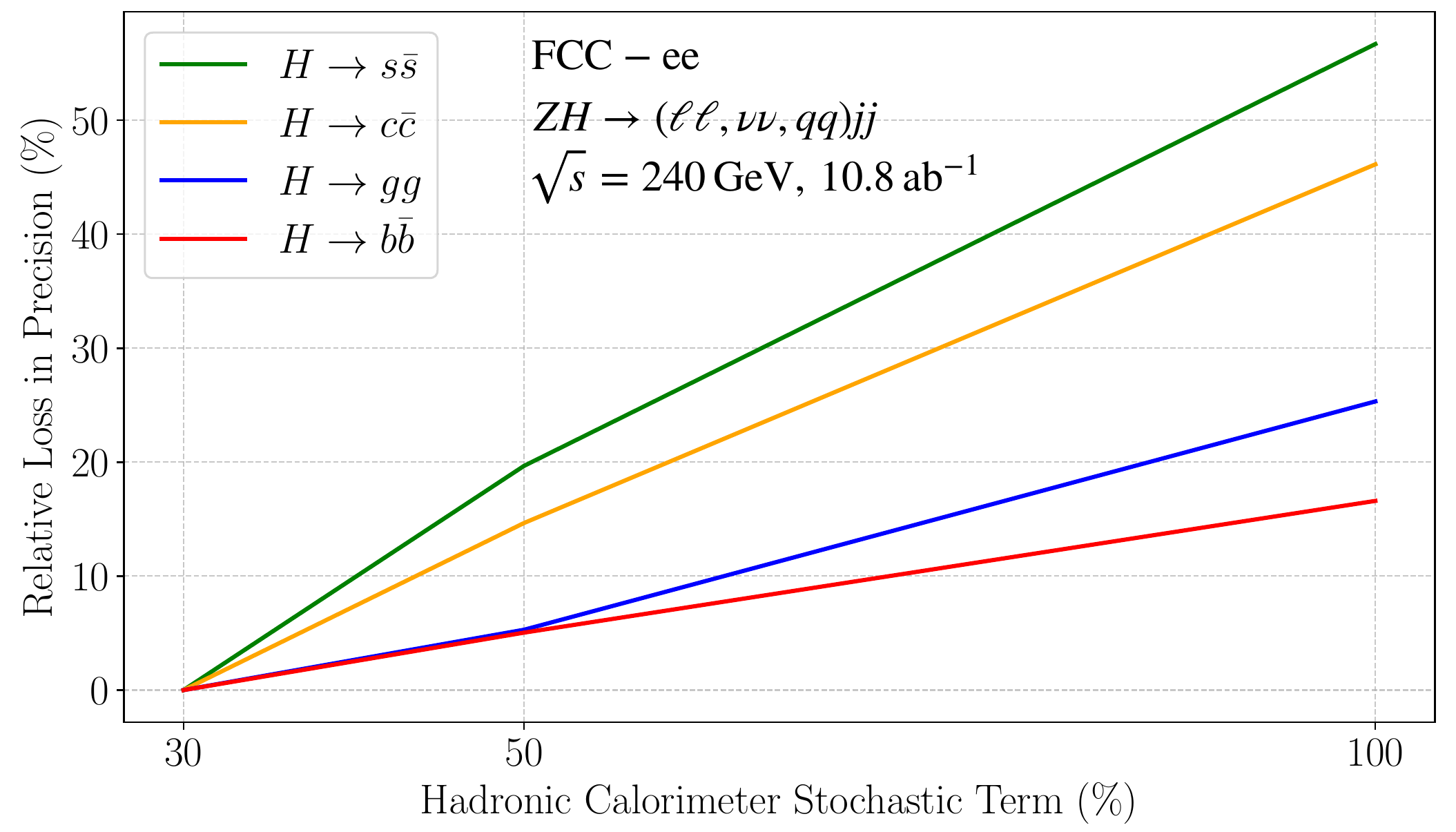}
    \caption{Expected precision degradation of the branching ratio measurements for the $H\to c\bar{c},b\bar{b},s\bar{s},gg$ decays
as a function of the HCAL stochastic term. To guide the interpretation, the value 30\% would correspond to a
dual-readout calorimeter as in the baseline IDEA simulation, 50\% corresponds to an ATLAS-type calorimeter, and
100\% to a CMS-type calorimeter.}
    \label{fig:caloVariations}
\end{figure}

The identification of di-jet signal events relies on mass resolution (the visible mass), which is influenced by calorimeter performance. Assuming an ideal particle-flow algorithm, visible energy (and mass) resolution is dominated by the neutral hadron resolution, which is primarily driven by the HCAL stochastic term. Figure~\ref{fig:caloVariations} shows the relative loss in precision in $H\to c\bar{c},b\bar{b},s\bar{s},gg$ decay channels as a function of the HCAL stochastic term which shows that the visible energy (and mass) resolution plays significant role in the branching fraction measurements especially for $H\to c\bar{c}$ and $H\to s\bar{s}$ decay channels.

Additional studies with various tracker assumptions were performed by increasing the beam pipe material, displacing the vertex detector by $500\,\mu \mathrm{m}$, or varying its resolution by small amounts; however, no significant effect on precision loss was observed.
 
\section{Higgs self-coupling interpretation}
\label{sec:couplings}
The Higgs self-coupling is a parameter of major interest, as it has a direct impact on the shape of the Higgs potential. It also has a predicted value in the SM, determined by the Higgs mass and vacuum expectation value, to which measurements at colliders can be compared. This parameter is directly accessible through Higgs pair production (HH), a process actively studied at the LHC. It is possible to produce HH at an $e^{+}e^{-}$ collider and access the Higgs self-coupling via the Higgs-strahlung process, provided that the $e^{+}e^{-}$ center-of-mass energy is large enough. At the 240 GeV phase of FCC-ee, $\sqrt{s}$ is too low to produce HH. Instead, the Higgs self-coupling can be indirectly accessed through higher order contributions to single Higgs production processes. In general for $e^{+}e^{-}$ colliders, the cross-section for HH starts to become experimentally viable around $\sqrt{s} = 500$ GeV. Current feasibility studies estimate that with $\sqrt{s} = 240$ GeV, a sensitivity to the self-coupling of 49\% may be achievable. The recent FCC Feasibility Study reports an expected precision on the Higgs self-coupling of about 28\%, including input from the $\sqrt{s} = 240$ GeV and $\sqrt{s} = 365$ GeV runs. When combined with the recent HL-LHC (High Luminosity LHC) projections~\cite{atlas2025highlightshllhcphysicsprojections}, the precision on the self-coupling is expected to reach 18\% \cite{Benedikt:2928193}.

\subsection{Theoretical framework}

BSM contributions to the ZH cross-section can be parameterized within the SMEFT (SM Effective Field Theory) framework. In our analysis, we use SMEFT in the Warsaw basis~\cite{Warsaw}, and focus on the contribution of the $C_{\phi}$~\cite{SMEFT_1}~\cite{SMEFT_2} operator, which affects the Higgs self-coupling via a dimension-6 operator.

The following equation is used to parameterize the ZH cross-section as a function of $C_{\phi}$:

\begin{equation}\label{eq:Cphi}
    \frac{\sigma_{\rm SMEFT}(C_\phi,\sqrt{s}=240 \, \rm GeV, \Lambda = 1 \, \rm TeV)}{\sigma_{\rm SM,LO}} = 1 - 0.00699C_{\phi}(\mu)
\end{equation}
 
It is important to note that for single Higgs production processes, contributions of $C_{\phi}$ for the ZH process start at the one-loop level. Therefore, this contribution depends on the renormalization scheme choice. In this study, we follow~\cite{SMEFT_1}~\cite{SMEFT_2} where $\mu$ is evaluated at 240 GeV. Additionally, in this study, only the cross-section of the ZH process is parameterized, while the branching ratios are left at their SM values.

\subsection{Analysis methodology}

As described in Sec.~\ref{sec:ZHfullHad}, the fully-hadronic ZH analysis can be used to study various Higgs couplings, and has been performed in~\cite{GeorgesPaper1}. For the interpretation described in this section, the same events passing the selections and categorization of the nominal fully-hadronic ZH analysis described in~\cite{GeorgesPaper1} are used, with the ZH cross-section parameterized as a function of $C_{\phi}$ using Eq.~\ref{eq:Cphi}. 

This parameterization changes the cross-section in a linear fashion, up to about 0.7\%, when varying $C_{\phi}$ from -1 to 1. This range of $C_{\phi}$ is considered accurate without including higher order terms in its expansion, as including higher powers of $C_{\phi}$ returns almost the same cross-section variation in this range. Additionally, setting $C_{\phi}$ to 0 leaves the cross-section unaltered, at its SM value. 

\subsection{Results and detector sensitivity}

\begin{figure}[hbtp]
    \begin{center}
        \includegraphics[width=0.82\columnwidth]{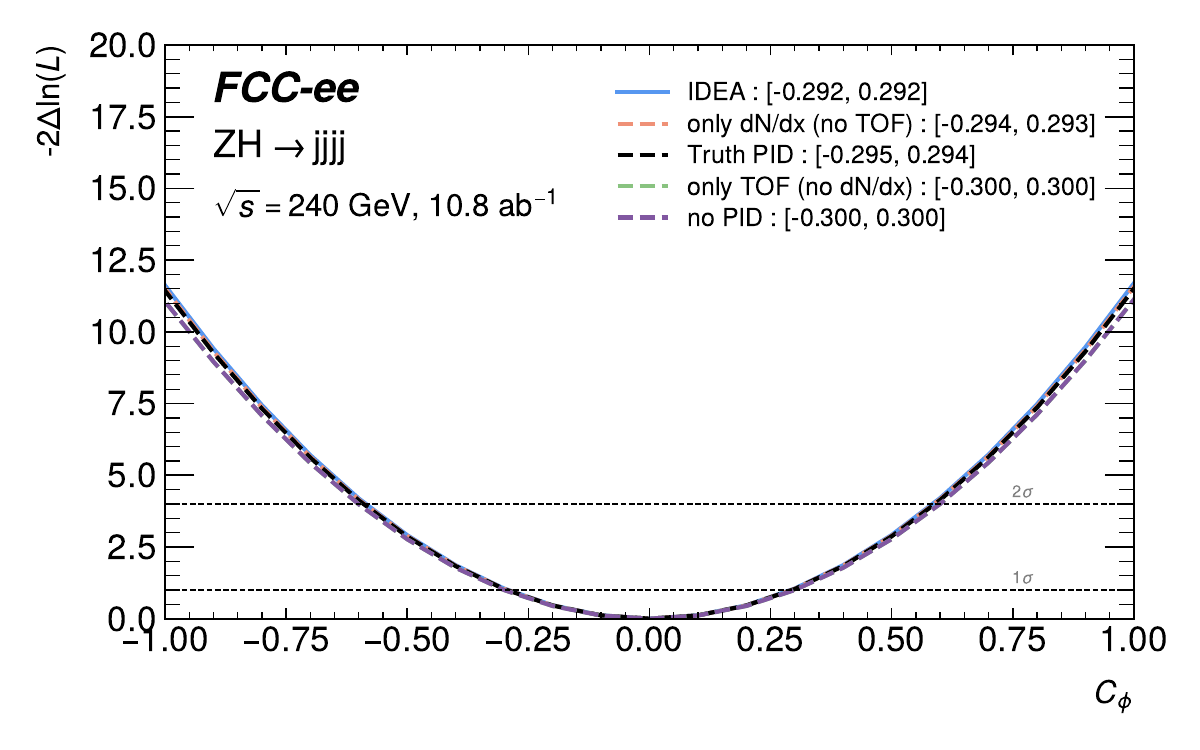}
        \caption{Negative log-likelihood scan as a function of the SMEFT parameter $C_{\phi}$. We compare the baseline IDEA detector tagger with variants trained without time-of-flight information (no TOF, only dN/dx), without cluster counting (no dN/dx, only TOF), without both dN/dx and TOF (no PID), and with truth-level particle identification (Truth PID). The horizontal lines indicate the 68\% and 95\% confidence levels, representing one and two standard deviation uncertainties, respectively.}
        \label{fig:selfcouplingscans}
    \end{center}
\end{figure}

We apply this parameterization to all ZH processes and perform likelihood scans for the three detector variations described in Section~\ref{sec:impactofTaggerPerf}, the ``No PID" scenario, and the ``Truth PID" scenario. The result is shown in Fig.~\ref{fig:selfcouplingscans}.

The constraints on $C_{\phi}$ show very similar behavior to the other Higgs coupling scans shown in Fig.~\ref{fig:ZHScans}. This consistency is expected since the same events are used for this interpretation as for the coupling scans shown in Fig.~\ref{fig:ZHScans}, and the variation of the cross-section from $C_{\phi}$ is only up to about 0.7\%. A recent $C_{\phi}$ projection including a combination with HL-LHC projections can be found in \cite{terhoeve2025higgstrilinearcouplingsmeft}.

The observed consistency in $C_{\phi}$ constraints across different detector variations aligns with our expectations. The detector variations described in Section~\ref{sec:impactofTaggerPerf} predominantly impact the coupling of the Higgs to strange quarks, as shown in Fig.~\ref{fig:HssScan}. For the Higgs self-coupling interpretation, all categories are included to maximize sensitivity, among which the Higgs to strange quarks category is sub-dominant and therefore does not strongly influence the combined $C_{\phi}$ constraint across different detector variations.

\section{Higgs-to-invisible analysis}
\label{sec:invisible}

The invisible decay of the Higgs boson represents an important probe for physics beyond the Standard Model, with numerous theoretical models predicting contributions beyond the SM process $H \rightarrow ZZ \rightarrow 4\nu$. The clean environment of electron-positron collisions enables reconstruction of unobserved Higgs boson decays through energy-momentum conservation and knowledge of the initial state. Thus, the precision of this measurement is highly sensitive to detector design and simulation accuracy.

We explore this sensitivity through two complementary approaches. First, in Sect. \ref{sec:hinv-selection}--\ref{sec:hinv-fullsim-limits}, we update the previous analysis~\cite{andy-hinv} that relied on fast simulation (for the IDEA detector~\cite{IDEA1,IDEA2}) to incorporate full simulation (for the CLD detector~\cite{cld}) for the signal and primary backgrounds, enabling direct comparison between fast and full simulation methodologies. Second, in Section \ref{sec:hinv-detector-variation}, we use fast simulation of the IDEA detector to quantify how the calorimeter property variations described in Section \ref{sec:calo} affect measurement sensitivity.

\subsection{Selection and analysis}
\label{sec:hinv-selection}

\begin{figure}[hbtp]
  \begin{center}
    \includegraphics[width=0.62\columnwidth]{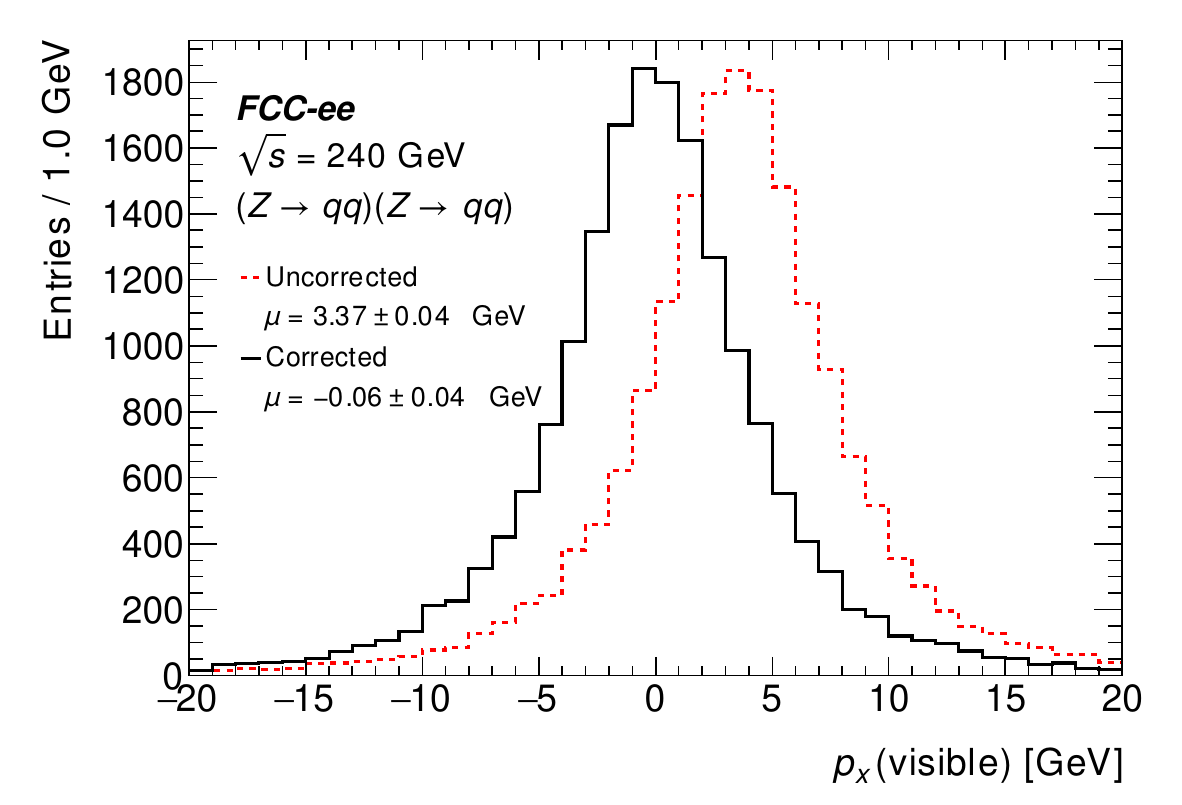}
    \caption{Total momentum of final-state particles in the $x$-direction for $ZZ \rightarrow qqqq$ events. The dashed curve shows the distribution before applying the compensating boost, while the solid curve shows the distribution after correction.}
    \label{fig:vispx}
  \end{center}
\end{figure}

We base our selection on the previous fast-simulation analysis~\cite{andy-hinv}, adapted for full simulation. The procedure begins by boosting all final-state particles into the center-of-mass frame as described below.
At the FCC, each beam enters the detector at an angle of 0.015 rad with respect to the $z$-axis in the horizontal plane, resulting in a total crossing angle of 0.030 rad. This configuration imparts a boost to events in the $x$-direction of $\beta=\sin(\theta/2)$. Figure \ref{fig:vispx} shows that the total momentum in the $x$-direction for $ZZ \rightarrow qqqq$ events is offset from zero by several~GeV. Since this offset would affect missing momentum reconstruction, we apply a compensating boost in the $x$-direction to all final-state particles in fully-simulated samples.

Since this effect is absent in fast simulation, no correction was applied to those samples.

We select electrons, muons, and photons using the tight criteria from \progname{PandoraPFA} \cite{Thomson_2009}, requiring leptons to have momentum $p > 10$ GeV and satisfy isolation criteria. For isolation, we calculate the scalar sum $p_\textrm{cone}$ of momenta for all particles within a cone of $0.01 < \Delta R < 0.5$ around each lepton, where $\Delta R = \sqrt{(\Delta\phi)^2 + (\Delta\eta)^2}$ and $\eta$ is pseudorapidity. A lepton is considered isolated if $p_{\ell} / p_{\textrm{cone}} > 0.5$.

We then apply a bremsstrahlung recovery procedure described below.

\begin{figure}[!htbp]
  \begin{center}
    \subfloat[]
    {\includegraphics[width=0.49\textwidth]{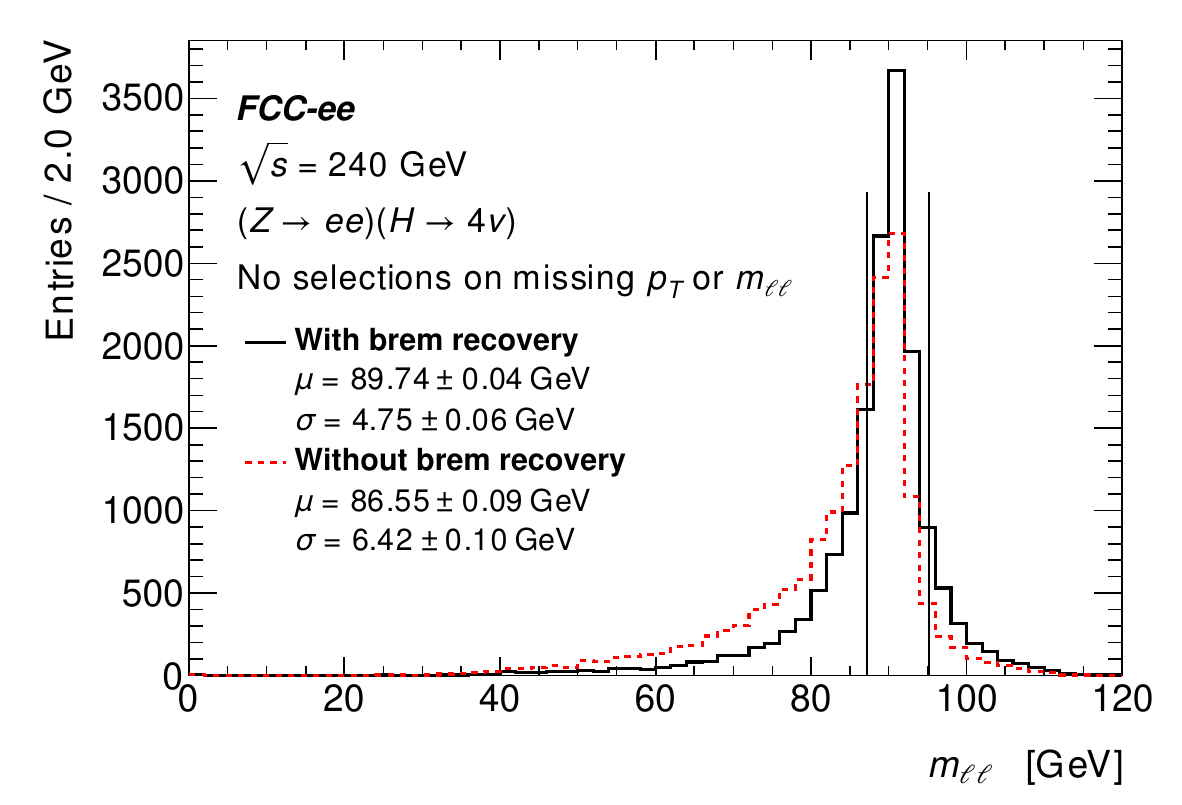}\label{fig:hinv-brem-ee}
    }
    \subfloat[]
    {\includegraphics[width=0.49\textwidth]{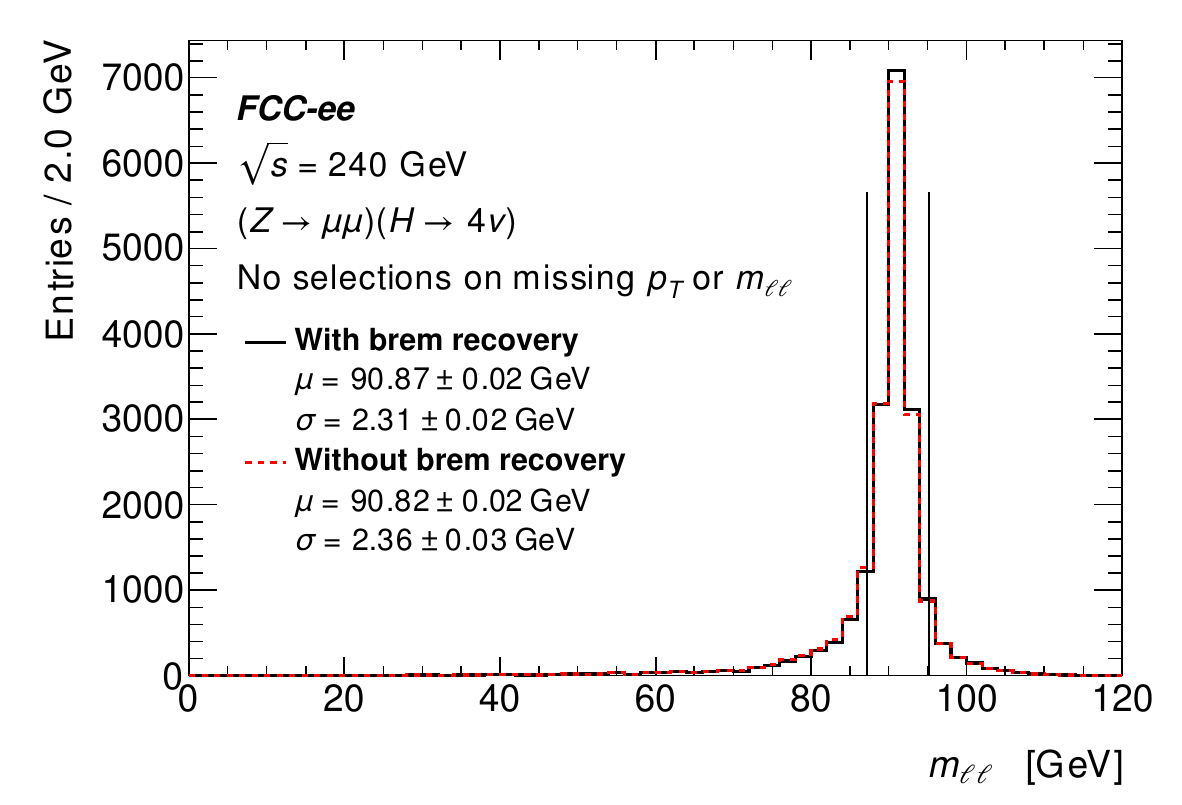}\label{fig:hinv-brem-mm}
    }
    \caption{Dilepton invariant mass for $ZH$ events with the Higgs boson decaying to four neutrinos and the $Z$ boson decaying to (a) electrons and (b) muons. The dashed curves show distributions without bremsstrahlung correction, while solid curves show distributions after applying the correction. Parameters of Gaussian functions fit to the distributions are shown. Vertical lines indicate the acceptance range for analysis selection.
      \label{fig:hinv-brem}
    }
  \end{center}
\end{figure}

Electrons traversing detector material emit photons at small angles (bremsstrahlung), producing a substantial low-mass tail in the $m_{ee}$ distribution, as shown by the dashed curve in Figure \ref{fig:hinv-brem-ee}. Since this effect is neither included in the fast simulation nor accounted for in standard CLD reconstruction, it represents a significant source of discrepancy between fast and full simulation. To reduce these differences, we implemented a bremsstrahlung recovery procedure at the analysis level.

Since refitting tracks at the analysis level is not feasible, we instead identify the highest-momentum photon candidate within a 0.05 rad cone around each lepton, then determine whether to add this photon's momentum to the lepton. Simply adding all nearby photons to leptons would degrade resolution, as photon momenta may already be incorporated into lepton tracks depending on where radiation occurred in the detector. Therefore, we trained a neural network (NN) to identify cases where photon addition improves momentum reconstruction.

We classify leptons based on whether adding the photon brings the lepton's momentum closer to or further from the momentum given by MC truth information. Photons that improve momentum reconstruction are classified as ``signal'' while those that worsen it were labeled ``background''. The NN processes five input variables: lepton momentum $p_{\ell}$, the lepton polar angle $\theta_{\ell}$, the angle between the lepton and photon, $\Delta\theta(\ell,\gamma)$, the energy of the lepton's calorimeter cluster $E_\ell^{\text{calo}}$, and the energy of the photon's calorimeter cluster $E_\gamma^{\text{calo}}$. The NN is implemented using the DNN method from TMVA~\cite{TMVA} with two hidden layers containing 14 and 6 nodes, using $\tanh$ activation functions for hidden layers and a linear activation function for the output node. The network is trained on the $(Z \rightarrow ee)(H \rightarrow 4\nu)$ signal sample, equally divided between training and test samples. For photons with NN output greater than 0.5, their momentum is added to the associated lepton.

This correction substantially improves the electron pair mass distribution, as shown by the solid curve in Figure \ref{fig:hinv-brem-ee}. The distribution becomes narrower and more symmetric, with the mean shifting closer to the $Z$ boson mass. We apply the same NN to muons, which, despite being less affected by bremsstrahlung, still show slight narrowing of the mass distribution (Figure \ref{fig:hinv-brem-mm}).

The bremsstrahlung recovery procedure is followed by calculation of the total visible momentum $p_{\textrm{vis}}$ by summing over all final-state particles, excluding photons that the NN determined should not be added to leptons. For backgrounds simulated with fast simulation, we omit the center-of-mass boost and bremsstrahlung recovery steps, and select leptons using the provided identification lists.

The analysis divides into three channels based on $Z$ boson decay: electrons, muons, and hadrons. For leptonic channels, we require exactly two same-flavor, opposite-sign leptons, transverse missing momentum $p_{T,\textrm{vis}} > 10$ GeV, and dilepton mass within 4 GeV of the $Z$ mass ($|m_{\ell\ell} - M_Z| < 4$ GeV). For the hadronic channel, we require no identified leptons, transverse missing momentum $p_{T,\textrm{vis}} > 15$ GeV, and visible mass between 86 and 105 GeV. Table \ref{tab:hinv-selection} summarizes these selection criteria.

\begin{table}
  \centering
  \caption{Event selection criteria for the invisibly-decaying Higgs boson analysis using full simulation.}
  \label{tab:hinv-selection}
  \begin{tabular}{|l|>{\centering}m{5cm}|c|}
    \hline
    Leptons
      & \multicolumn{2}{>{\centering}m{10cm}|}{Tight electrons or muons with $p>10$ GeV \\
     Apply bremsstrahlung recovery} \\
    \hline
    Channel definition
      & \multicolumn{1}{>{\centering}m{5cm}|}{Exactly two leptons:  \\
                                $e^+e^-$ or $\mu^+\mu^-$}
      & No leptons \\
    \hline
      Missing $p_T$ & $p_{T,\textrm{vis}} > 10$ GeV & $p_{T,\textrm{vis}} > 15$ GeV \\
    \hline
    $Z$ boson decay & $|m_{\ell\ell} - M_Z| < 4$ GeV & $86$ GeV $< m_\textrm{vis} < 105$ GeV \\
    \hline
  \end{tabular}
\end{table}

\begin{figure}[hbtp]
  \begin{center}
    \includegraphics[width=0.45\columnwidth]{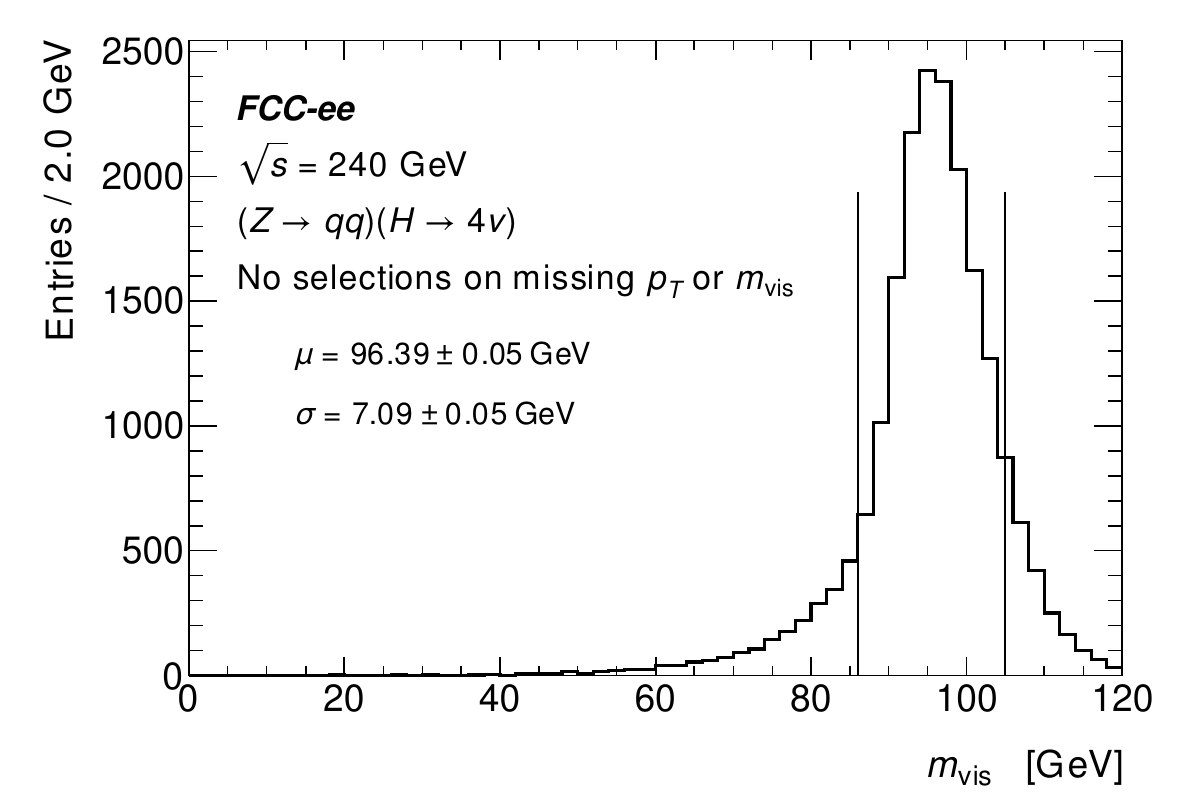}
    \caption{Invariant mass distribution of all visible particles for $ZH$ events with the Higgs boson decaying to four neutrinos and the $Z$ boson decaying hadronically. The mean and width from a Gaussian fit to the distribution are shown. Vertical lines indicate the acceptance range for analysis selection.}
    \label{fig:hinv-mvis-qq}
  \end{center}
\end{figure}

For the hadronic channel, we expanded the visible mass selection range compared to the previous fast-simulation analysis~\cite{andy-hinv}. Figure \ref{fig:hinv-mvis-qq} shows the visible mass distribution for the hadronic channel, which peaks significantly above the $Z$ boson mass. This shift likely results from particle flow reconstruction artifacts in \progname{PandoraPFA}, where neutral clusters near charged hadrons are reconstructed as separate photons despite their momentum being included in the charged hadron's track, effectively double-counting energy in the particle summation.

\begin{figure}[!htbp]
  \begin{center}
    \subfloat[]{\includegraphics[width=0.35\textwidth]{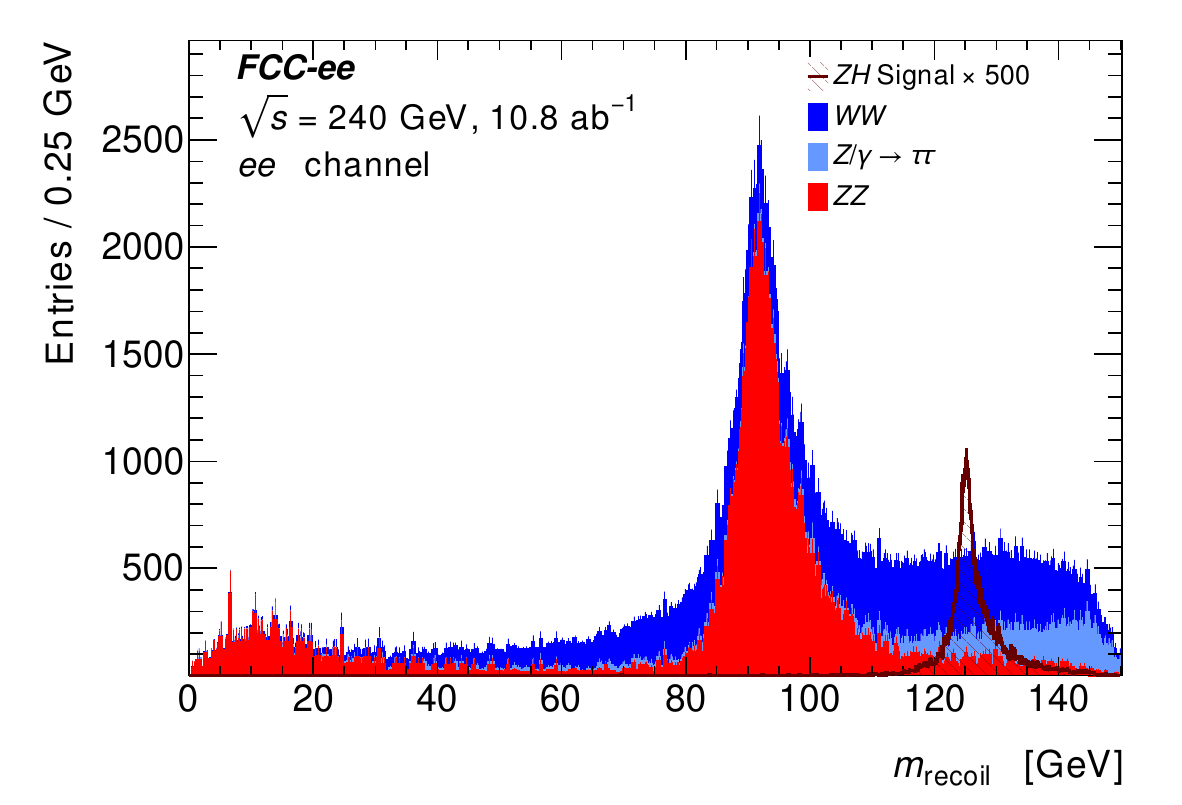}}
    \subfloat[]{\includegraphics[width=0.35\textwidth]{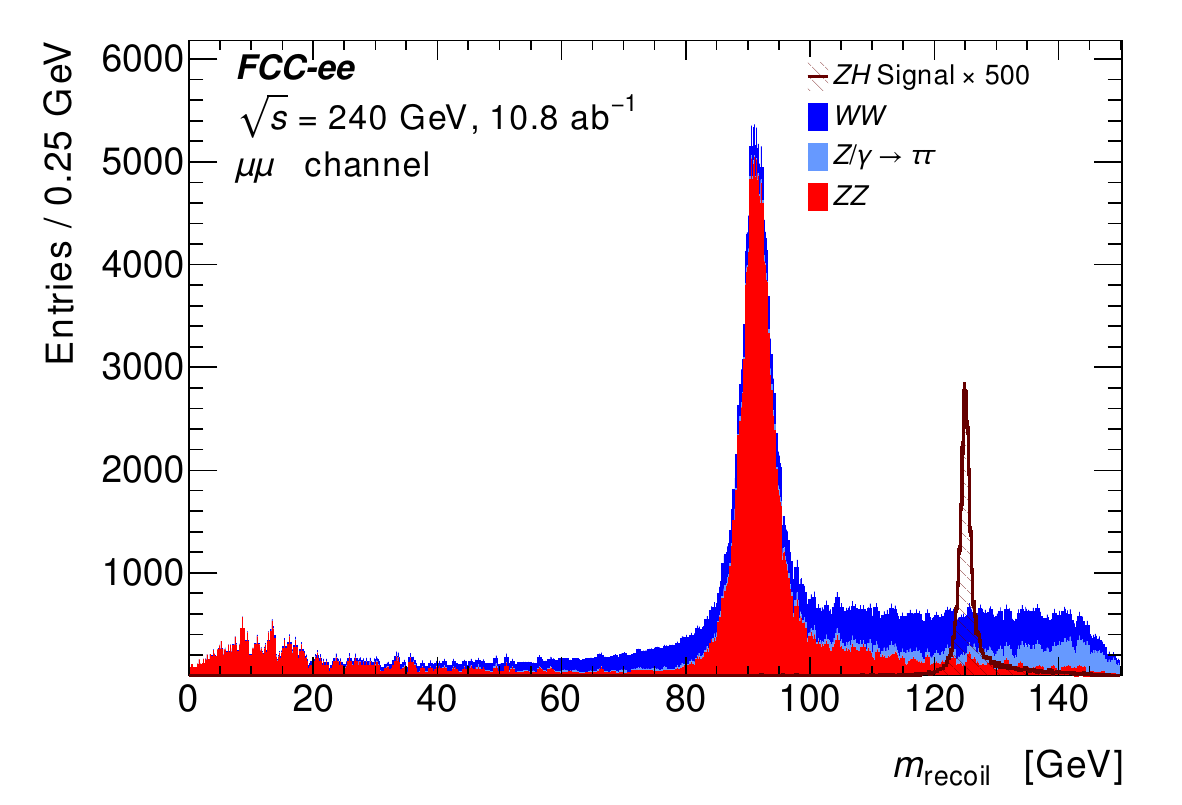}}
    \subfloat[]{\includegraphics[width=0.35\textwidth]{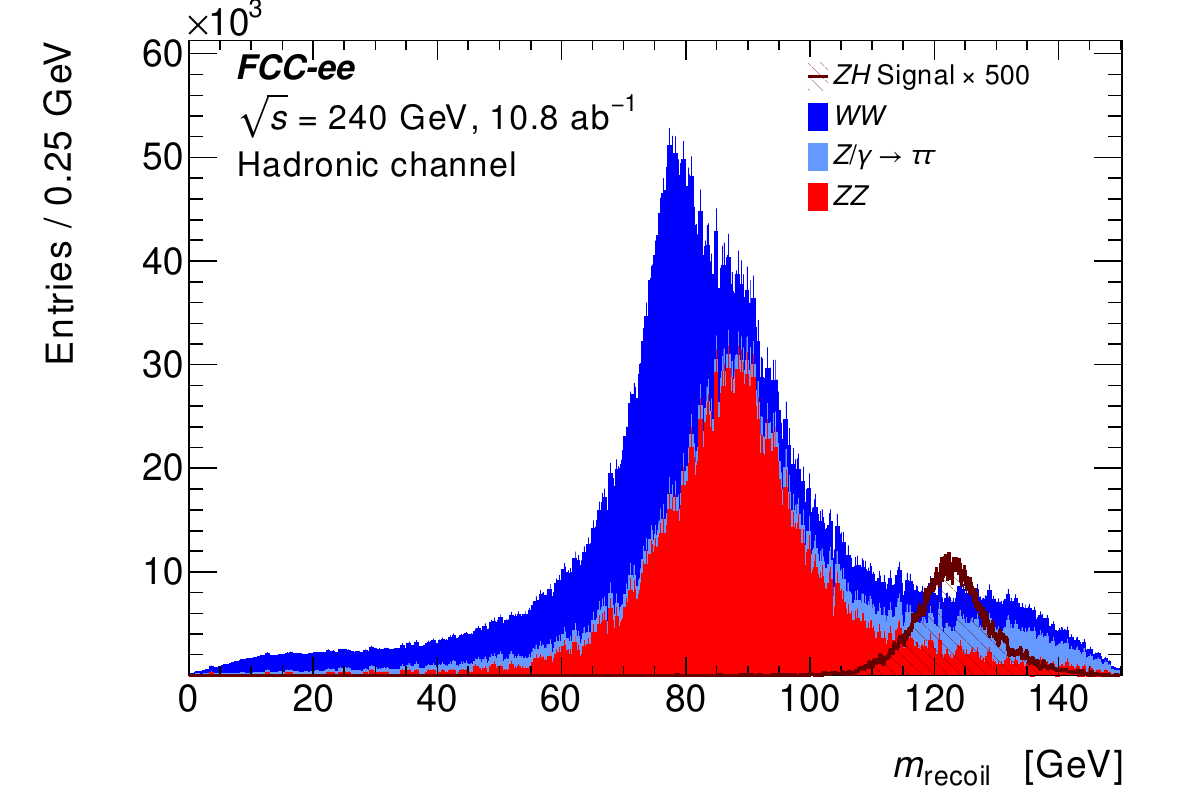}}
    \caption{Reconstructed invisible recoil mass distributions after all selections, normalized to an integrated luminosity of 10.8 ab$^{-1}$ for the (a) $ee$, (b) $\mu\mu$, and (c) hadronic channels.
      The signal (red) cross section is scaled up by a factor of~500.
      The background from other $ZH$ decay modes is negligible and
      not included in these plots.}
    \label{fig:recoilMass}
  \end{center}
\end{figure}

Finally, we calculate the recoil mass from the total visible momentum and nominal collision energy. Figure \ref{fig:recoilMass} shows the resulting distributions for all three channels.

\subsection{Resolution comparison between fast and full simulation}
\label{sec:resolution-comparison}

\begin{figure}[hbtp]
  \begin{center}
    \subfloat[]
    {\includegraphics[width=0.49\textwidth]{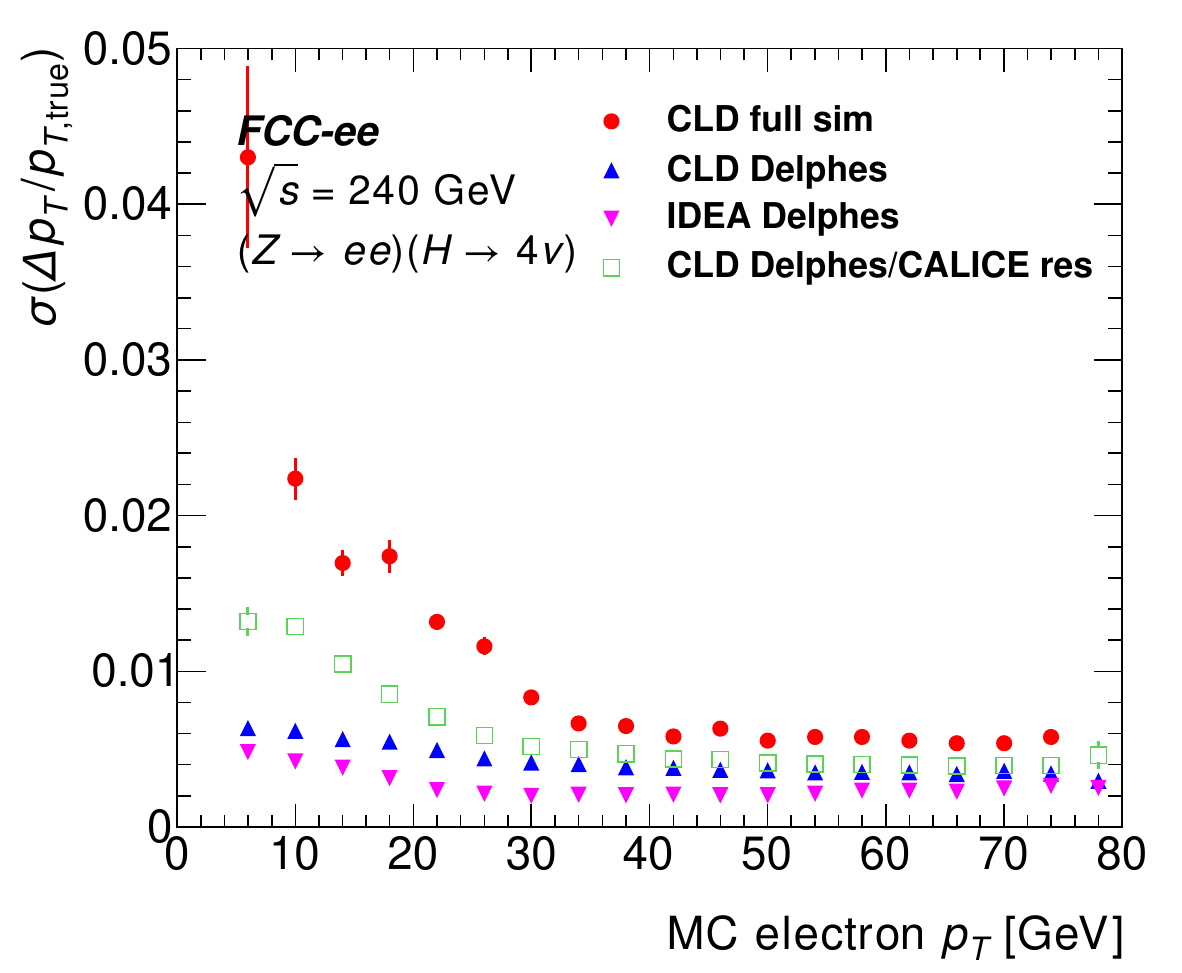}\label{fig:electron-resolution}
    }
    \subfloat[]
    {\includegraphics[width=0.49\textwidth]{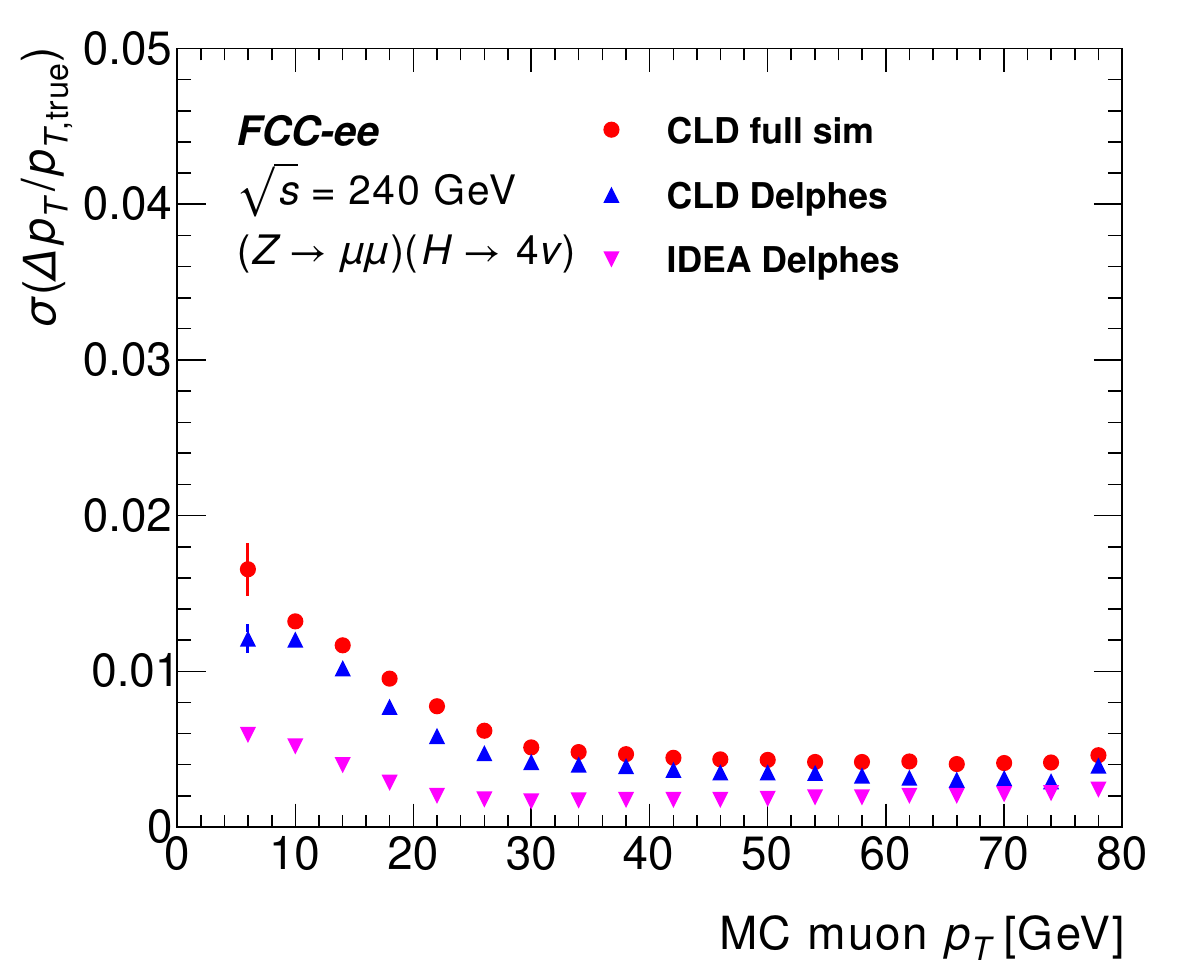}}
    \caption{Relative transverse momentum resolution as a function of transverse momentum for (a) electrons and (b) muons. Within each momentum bin, resolution distributions are fit to Gaussian functions. Results are shown for full simulation (without bremsstrahlung recovery), fast simulation for both CLD and IDEA configurations, and modified CLD fast simulation using a silicon-tungsten CALICE-like \cite{CALICE:2012ami} calorimeter resolution function.}
    \label{fig:ElectronMuonResolution}
  \end{center}
\end{figure}

Figure \ref{fig:ElectronMuonResolution} compares the transverse momentum resolution for electrons and muons between full simulation and fast simulation for both CLD and IDEA detector configurations. For muons, whose momentum resolution is determined almost entirely by tracking performance, the CLD fast simulation closely matches the full simulation results. The IDEA fast simulation shows significantly better muon momentum resolution than CLD, particularly at low momenta, as expected from the lower material budget of its drift chamber-based tracking compared to CLD's all-silicon tracker.

For electrons, however, the CLD fast simulation significantly outperforms full simulation, showing even better resolution than for muons at low momenta. This discrepancy arises because the electromagnetic calorimeter resolution in both CLD and IDEA fast simulation configurations corresponds to a crystal calorimeter ($\sigma_E/E = 0.5\% \oplus 3\%/\sqrt{E} \oplus 0.2\%/E$), while CLD actually employs a silicon-tungsten CALICE-style calorimeter with expected resolution $\sigma_E/E = 1\% \oplus 16\%/\sqrt{E}$~\cite{2008.00338}. When we modify the CLD fast simulation to use the CALICE-like calorimeter resolution (Figure \ref{fig:electron-resolution}), the results align much more closely with full simulation. This finding highlights the need for caution when using existing fast simulation for applications where electromagnetic calorimeter resolution is critical.

\subsection{Results}
\label{sec:hinv-fullsim-limits}

To determine upper limits on the branching ratio ($\text{BR}_{\text{inv}}$) of the Higgs boson decaying to invisible final states, we use the statistical methodology explained in Sec.~\ref{sec:stats}.
Table \ref{tab:limitTable} presents the resulting limits for each channel, representing an idealized scenario with statistical uncertainties only. The hadronic channel provides the strongest constraint, with an expected limit of $2.8 \times 10^{-3}$ at 95\% CL.

\begin{table}
  \centering
  \caption{Expected 95\% CL upper limits on the Higgs-to-invisible branching ratio by channel.}
  \label{tab:limitTable}
  \begin{tabular}{|c|c|c|c|c|c|}
    \hline
    Channel & $-2\sigma$ & $-1\sigma$ & Median & $+1\sigma$ & $+2\sigma$ \\
    \hline
    $ee$ & 6.0$\times10^{-3}$&8.1$\times10^{-3}$&1.1$\times10^{-2}$&1.6$\times10^{-2}$&2.1$\times10^{-2}$ \\
    \hline
    $\mu\mu$ & 2.5$\times10^{-3}$&3.4$\times10^{-3}$&4.7$\times10^{-3}$&6.6$\times10^{-3}$&8.9$\times10^{-3}$ \\
    \hline
    $qq$ & 3.6$\times10^{-3}$&4.8$\times10^{-3}$&6.7$\times10^{-3}$&9.3$\times10^{-3}$&1.3$\times10^{-2}$ \\
    \hline
  \end{tabular}
\end{table}

\subsection{Detector variations}
\label{sec:hinv-detector-variation}

\begin{figure}[!htbp]
  \begin{center}
    \includegraphics[width=0.8\textwidth]{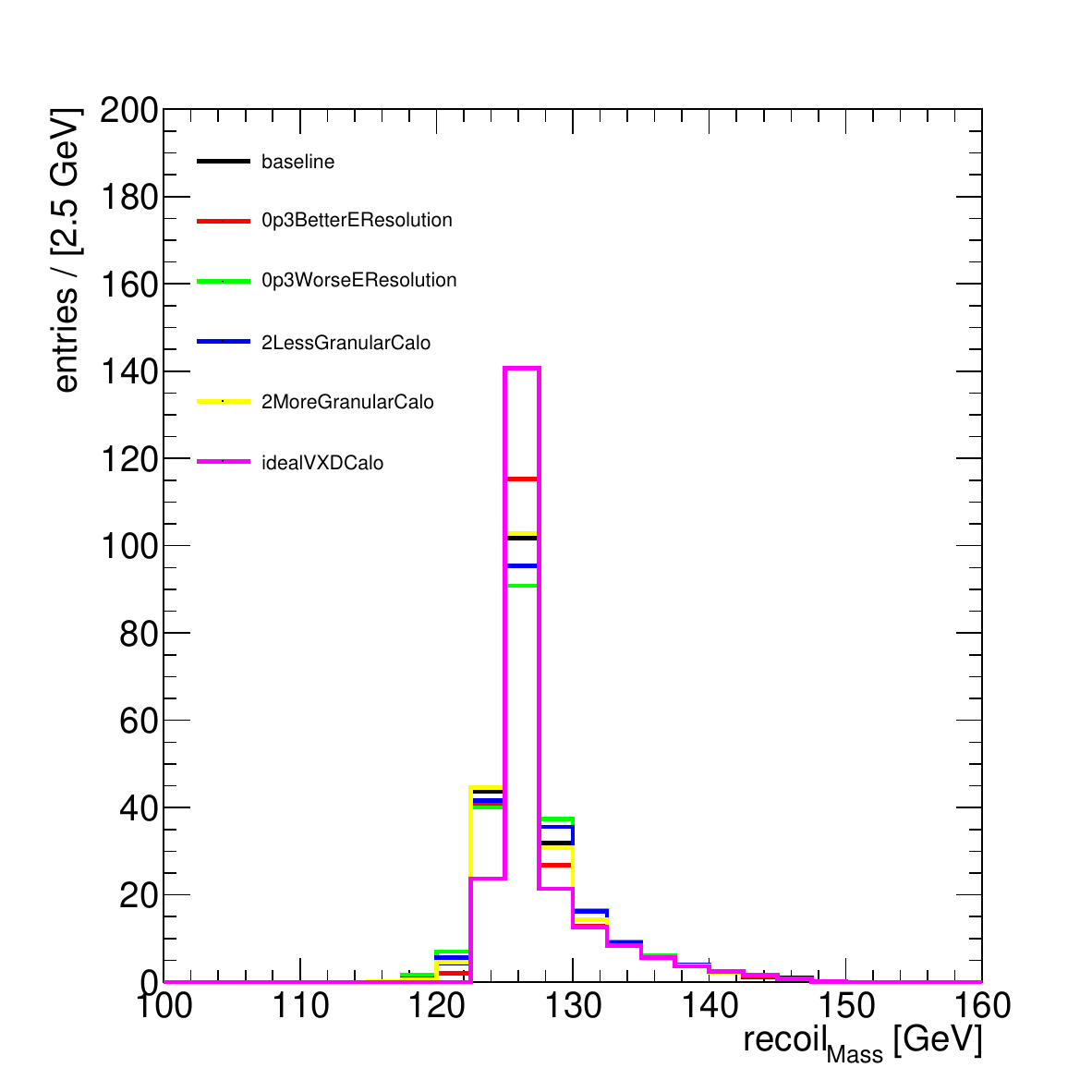}
    \caption{Reconstructed invisible recoil mass distributions for $ZH$ signal in the hadronic channel, normalized to corresponding cross sections. Comparison includes baseline IDEA calorimeter (black), 30\% better relative energy resolution (red), 30\% worse relative energy resolution (green), two times more granular calorimeter (yellow), and two times less granular calorimeter (blue).}
    \label{fig:recoilMass-detvar}
  \end{center}
\end{figure}

We propagated variations in calorimeter properties, as described in Section \ref{sec:calo}, through the Higgs-to-invisible analysis \cite{andy-hinv}. Figure \ref{fig:recoilMass-detvar} shows the reconstructed invisible recoil mass distributions normalized to corresponding cross sections for the $ZH$ signal in the hadronic channel.

To assess the impact of these detector configurations on sensitivity, we determined the limit on the Higgs-to-invisible branching ratio for each variation. We generated six detector configurations using fast simulation, corresponding to the signal distributions in Figure \ref{fig:recoilMass-detvar}, while using the nominal detector configuration for backgrounds. We focused exclusively on the hadronic channel due to its superior sensitivity (Table \ref{tab:limitTable}).

Table \ref{tab:limitTable-detvar} summarizes the relative changes in expected limits compared to the baseline configuration. Notably, improved energy resolution enhances sensitivity by approximately 9\%, while degraded resolution reduces sensitivity by 11\%. Increased calorimeter granularity has minimal effect, but reduced granularity degrades sensitivity by 7\%. The combined effect of an ideal vertex detector and calorimeter improves sensitivity by approximately 27\%, demonstrating the potential benefit of detector optimization for this crucial measurement.

\begin{table}
  \centering
  \caption{Relative change in expected 95\% CL upper limits on the Higgs-to-invisible branching ratio for different detector configurations compared to baseline (hadronic channel only). Positive values indicate improved sensitivity (lower limits), while negative values indicate decreased sensitivity.}
  \label{tab:limitTable-detvar}
  \begin{tabular}{|c|r|r|r|r|r|}
    \hline
    Detector configuration & $-2\sigma$ & $-1\sigma$ & Median & $+1\sigma$ & $+2\sigma$ \\
    \hline
    30\% Better E-Resolution & $+8.3\%$ & $+12.1\%$ & $+8.9\%$ & $+9.5\%$ & $+10.6\%$ \\
    \hline
    30\% Worse E-Resolution & $-12.5\%$ & $-9.1\%$ & $-11.1\%$ & $-11.1\%$ & $-10.6\%$ \\
    \hline
    2× Less Granular Calo & $-8.3\%$ & $-6.1\%$ & $-6.7\%$ & $-6.4\%$ & $-5.9\%$ \\
    \hline
    2× More Granular Calo & $0.0\%$ & $+3.0\%$ & $0.0\%$ & $0.0\%$ & $+1.2\%$ \\
    \hline
    Ideal VXD+Calo & $+25.0\%$ & $+27.3\%$ & $+26.7\%$ & $+27.0\%$ & $+25.9\%$ \\
    \hline
  \end{tabular}
\end{table}
 
\section{Conclusion}

In this paper, we have presented a comprehensive study of detector configuration effects on three key physics analyses at the FCC-ee. This study systematically assesses the impact of detector design variations on physics measurements, offering key insights for optimizing future collider detectors.

We demonstrated that jet flavor identification performance—essential for precision Higgs physics—exhibits varying sensitivities to different detector parameters. In particular, we quantified how single-point resolution, material budget, silicon layer placement, and particle identification capabilities influence flavor-tagging performance.

When propagating these detector variations through the Higgs coupling measurements in the fully hadronic final state, we observed that the analysis exhibits remarkable robustness to most detector configuration changes. However, the precision of Higgs-strange quark coupling measurements depends critically on cluster-counting capabilities, with the expected uncertainty increasing by a factor of 1.6 when this information is removed from the flavor tagger training. Our investigation of alternative jet clustering algorithms further revealed that the anti-$k_t$ algorithm with radius parameter $R = 1.0$ provides better Higgs mass reconstruction than the Durham-$k_t$ algorithm traditionally used in FCC-ee studies.

For the first time, we have also employed full detector simulation in the study of Higgs-to-invisible decays, enabling direct comparison with fast simulation results and identification of critical detector properties. We found that electromagnetic calorimeter resolution has a substantial impact on this analysis, with a 30\% improvement in energy resolution enhancing the branching ratio sensitivity by approximately 9\%. Similarly, calorimeter granularity affects measurement precision, with a factor of 2 reduction in granularity degrading sensitivity by about 7\%.

The results presented in this work provide essential guidance for understanding the interdependence between detector design choices and physics performance at the FCC-ee. By quantifying the impact of various detector configurations on key physics measurements, our findings will inform the optimization of detector designs for future experiments at electron-positron colliders, helping to maximize their discovery potential and precision measurement capabilities.
\clearpage

\section*{Acknowledgements}
We would like to acknowledge Sally Dawson for useful discussions and advice on the SMEFT interpretation.

This research was supported in part by the U.S. Department of Energy’s Office of Science, Office of High Energy Physics under contract no. DE-SC0012704.
This work was supported by resources provided by the Scientific Data and Computing Center (SDCC),  component of the Computational Science Initiative (CSI) at Brookhaven National Laboratory (BNL).

\bibliography{apssamp}

\end{document}